\documentclass[twocolumn]{aastex631}

\usepackage{graphicx}   
\usepackage{amsmath}    
\usepackage{mathrsfs}    
\usepackage{bm}
\usepackage{amssymb}    
\usepackage{diagbox}    
\usepackage{subfigure}
\usepackage{enumerate}
\usepackage{float}
\usepackage{ulem}
\usepackage{url}
\usepackage{soul}
\shorttitle{SN Ia from Pop III star}
\shortauthors{Z. Li et al.}

\begin{document}

\title{Type Ia Supernovae from First-generation Stars}

\correspondingauthor{Zhenwei Li; Lifan Wang}
\email{lizw@ynao.ac.cn; lifan@tamu.edu}

\author[0000-0002-1421-4427]{Zhenwei Li}
\affiliation{Yunnan Observatories, Chinese Academy of Sciences, Kunming, 650216, People's Republic of China}
\affiliation{Key Laboratory for the Structure and Evolution of Celestial Objects, Chinese Academy of Science, Kunming, 650216, People's Republic of China}
\affiliation{International Centre of Supernovae, Yunnan Key Laboratory, Kunming, 650216, People's Republic of China}


\author[0000-0001-7092-9374]{Lifan Wang}

\affiliation{George P. and Cynthia Woods Mitchell Institute for Fundamental Physics $\&$ Astronomy, Texas A$\&$M University, Department of Physics and Astronomy, 4242 TAMU, College Station, TX 77843, USA}

\author[0000-0001-9204-7778]{Zhanwen Han}
\affiliation{Yunnan Observatories, Chinese Academy of Sciences, Kunming, 650216, People's Republic of China}
\affiliation{Key Laboratory for the Structure and Evolution of Celestial Objects, Chinese Academy of Science, Kunming, 650216, People's Republic of China}
\affiliation{International Centre of Supernovae, Yunnan Key Laboratory, Kunming, 650216, People's Republic of China}

\author[0000-0001-5284-8001]{Xuefei Chen}
\affiliation{Yunnan Observatories, Chinese Academy of Sciences, Kunming, 650216, People's Republic of China}
\affiliation{Key Laboratory for the Structure and Evolution of Celestial Objects, Chinese Academy of Science, Kunming, 650216, People's Republic of China}
\affiliation{International Centre of Supernovae, Yunnan Key Laboratory, Kunming, 650216, People's Republic of China}





\begin{abstract}

  Type Ia Supernovae (SNe Ia) discovered at redshift $z\lesssim2.5$ are presumed to be produced from  Population (Pop) I/II stars. 
  {In this work, we investigate the production of SNe Ia from Pop III binaries in the cosmological framework. We derive the SN Ia rate as a function of redshift under a theoretical context for the production of first generation stars and emanate the likelihood of their detection by the James Webb Space Telescope (JWST).} 
  {Assuming the initial stellar mass function (IMF) favors low-mass stars as from recent numerical simulations, we found Pop III stars may give rise to a considerable amount of SNe Ia at high redshift and Pop III stars may even be the dominant SN Ia producer at z $\gtrsim 6$.} {In an optimistic scenario, we expect $\sim 1(2)$ SNe Ia from Pop III stars at $z\approx 4(5)$ for a survey of area $300 \;\rm arcmin^2$ during a $3\;\rm yr$ period with JWST. The same survey may record more than $\sim 400$ SNe Ia at lower redshift ($z\lesssim 2.5$) but with only about one of them from Pop III progenitors. There will be $\sim 6$ Pop III SNe Ia in the same field of view at redshifts of $5-10$.} 
  Observational constraints on SN Ia rates at the redshift range of $5-10$ can place crucial constraints on the IMF of Pop III stars. 
\end{abstract}




\section{Introduction}
\label{sec:1}

Type Ia supernovae (SNe Ia) are known as thermonuclear explosions of white dwarfs (WDs) with masses reaching the Chandrasekhar mass limit \citep{hoyle1960}. As a critical cosmic distance indicator, SNe Ia play an essential role in the discovery of the accelerating expansion of the Universe \citep{riess1998,schmidt1998,perlmutter1999}. The energetic explosion of SN Ia produces high energy cosmic-ray and heavy-elements which are important to galaxy evolution \citep{greggio1983,matteucci1986}. Currently, more than a few $1000$ SNe Ia have been observed out to $z\sim 2.5$ \citep[e.g.,][]{rodney2014,scolnic2018,des2024}, the event rates are related  to the cosmic star formation history (CSFH; \citealt{maoz2017}). The CSFH inferred from galaxy surveys in the UV and infrared suggests that most stars are born at $z\lesssim 4$ \citep{madau2014,madau2017}, which are generally classified as Population (Pop) I or II stars. An outstanding question is whether SNe Ia can arise from Pop III stars, i.e., the first-generation stars at much higher redshifts. 


Pop III stars are thought to be formed at {$z\sim 20-25$} and play a pivotal role in cosmic metal enrichment and re-ionization (see \citealt{klessen2023} for a review). These stars are produced from the collapse of metal-free primordial gas and evolve in distinct ways from the metal-rich (Pop I) and metal-poor (Pop II) stars \citep{marigo2001}. It has been argued that Pop III stars are formed with enormous mass (e.g., $\gtrsim 100M_\odot$; \citealt{omukai1998,abel2002,bromm2002,yoshida2008,hirano2014,hirano2015,hosokawa2016}), and the death of massive Pop III stars may lead to the formation of massive stellar black holes (BHs) due to the extremely weak or non-existence of wind mass-loss of Pop III stars \citep{heger2002,schaerer2002}. In particular, Pop III binaries may produce massive double BHs of which the gravitational wave (GW) will be detected by the third-generation ground-based GW interferometers, e.g., the Einstein Telescope \citep{punturo2010} and Cosmic Explorer \citep{reitze2019}. 

\citet{wangl2017} proposed the First Lights At REionization (FLARE) project to discover the distant SNe with James Webb Space Telescope (JWST), which will shed light on the properties of the SNe from the first stars. {These SNe may have both high and low mass progenitors and exhibit a broad range of delay time from the birth to the death of the stars. They are also greatly important for addressing many cosmological questions including the precise measurements of cosmological parameters.}
{
Unfortunately, no Pop III star has been confirmed in the observations so far (but see the possible Pop III signals found in \citealt{welch2022,wangx2022,maiolino2023}). Nevertheless, the supernovae from massive Pop III stars can potentially be observed out to high redshifts by the JWST \citep{gardner2006}. Core-collapse SNe (CCSNe), typically produced from $\sim 15-40\,M_\odot$ Pop III stars \citep{heger2002}, will be visible in the earliest galaxies out to $z \sim 10–15$ \citep{tanaka2013,whalen2013b}. Pair-instability SNe (PISNe) are generally more energetic than CCSNe, and capable of being observed at $z>25$ \citep{scanapieco2005,whalen2013c,whalen2014a,moriya2019}. The observations of primordial SNe could provide an opportunity to directly probe the mass spectrum of Pop III stars \citep{scanapieco2005,whalen2014b}. }


{
Recent works on the formation of Pop III stars suggest that the fragmentation phenomenon may lead to the birth of stars in the range of $\sim 1-10M_\odot$ (e.g., \citealt{turk2009,susa2014,stacy2016,wollenberg2020,sharda2020,jaura2022,prole2022,latif2022} and references therein). These stars may end their life as WD and then make it possible to produce SNe Ia. 
 }SNe Ia are mainly produced from two channels of progenitor systems, which are the single degenerate (SD) channel \citep{whelan1973,nomoto1984} and the double degenerate (DD) channel \citep{iben1984,webbink1984}. In the SD channel, the carbon-oxygen (CO) WD accretes mass from the non-degenerate companion, e.g., normal stars or helium (He) stars, resulting in thermonuclear runaway explosions as the mass of the WD increases to the maximum stable mass. In the DD channel, the SNe Ia are thought to arise from the mergers of two WDs in close orbit with a combined mass exceeding the Chandrasekher-mass limit. 
 These two models can account for many essential observational characteristics of SNe Ia (see the recent review of \citealt{liuz2023}). 


In this {work}, we aim to investigate the formation of SNe Ia from Pop III binaries. The recently proposed common envelope wind (CEW) model by \citet{meng2017} is adopted for the SD channel. The DD channel is also considered using a semi-analytic method. Then, we give the prediction of SNe Ia rate from Pop III binaries with the convolution of CSFH. The remainder of this paper is structured as follows. Section \ref{sec:2} presents the methods and results of binary evolution for SNe Ia. The calculation of SN Ia rate in the cosmological framework is introduced in Section \ref{sec:3}. The main results, including the prediction of SN Ia rate and implications for the observations, are presented in Section \ref{sec:4}. The summary and conclusion are given in Section \ref{sec:5}. 

\section{SN Ia from binary evolution}
\label{sec:2}

\subsection{Single Degenerate Channel}

\subsubsection{Methods and inputs}

In the SD scenario, SNe Ia are thought to be produced from the Chandrasekhar-mass WDs, which can naturally explain the homogeneity of most SNe Ia (see \citealt{liuz2023} for a review). The initial binary includes a CO WD and a non-degenerate star. Here we mainly focus on the {main sequence (MS)} star as the donor. The binary evolution simulations are performed via the detailed stellar evolution code {Modules for Experiments in Stellar Astrophysics} (\texttt{MESA}, Version 12115; \citealt{paxton2011,paxton2013,paxton2015,paxton2018,paxton2019}). 

{We first initialize the donor as a fully-convective star ($n=3$ polytropes) with the specified masses. This stage is known as the pre-MS stage. Then the pre-MS stars begin to contract under their own gravity until the nuclear burning produces enough energy to halt the contraction of the protostars. We choose the critical point when the radius of a protostar begins to expand as the start-point of zero-age MS (ZAMS) stage. In the subsequent evolution, we put the ZAMS models into the binary evolution modules. }

{The main \texttt{MESA} inputs in the binary evolution simulations are introduced as follows.} We adopt the Ledoux criterion and semi-convection mixing for the convection treatment, where the mixing length parameter is set to be $1.5$, and semi-convection is modeled with an efficiency parameter of $\alpha_{\rm sc}=0.01$. The convective overshooting is also considered for the convective cores and shells with the same parameters in \citet{herwig2000}. The Pop III stars born from the early epoch of the Universe are mainly composed of hydrogen (H), He, and {a little Li}. For stars with initial metallicity of $Z\leq 10^{-10}$, the CNO cycle cannot be activated initially \citep{cassisi1993}. Then we adopt the initial metallicity of $Z=10^{-14}$, the H mass fraction of $0.765$, and He mass fraction of $0.235$ \citep{lawlor2008,songh2020}. The stellar winds of Pop III stars should be drastically quenched due to the lack of heavy metal elements. Although some massive Pop III stars may still have non-negligible wind mass loss \citep{songh2020,aryan2023}, we are mainly concerned about the low- and intermediate-mass stars ($\lesssim 8M_\odot$) and the wind mass loss is ignored in the simulations. 

{The binary evolution simulations of Pop III stars have been studied in several previous works (e.g., \citealt{lawlor2008,songh2020,tsai2023}). In particular, \citet{tsai2023} performed the detailed evolutionary models of massive Pop III binaries with a large grid. Different from \citet{tsai2023}, we take the accretor as a point mass, and the mass transfer processes are simulated with the Kolb scheme \citep{kolb1990}}. {The accretion scenarios of the CO WDs are similar to many previous works} (e.g., \citealt{hachisu1996,han2004,meng2009,lizw2019}). The mass accumulation of a CO WD is strongly correlated with the critical accretion rate, which is defined by \citep{hachisu1999}
\begin{eqnarray}
  \frac{\dot{M}_{\rm cr}}{M_\odot\rm yr^{-1}} = 5.3\times 10^{-7}\frac{(1.7-X)}{X}\left(\frac{M_{\rm WD}}{M_{\odot}}-0.4\right),
  \label{eq:1}
\end{eqnarray}
where $X$ is the H mass fraction and $M_{\rm WD}$ is the WD mass. If the mass transfer rate, $\dot{M}_{\rm MT}$, is higher than $\dot{M}_{\rm cr}$, the accumulated material on the WD surface leads to the expansion of the WD. The WD will become a red giant-like object. The optically thick wind (OTW) model is often adopted for stars with high metallicity \citep{hachisu1996}, in which the accreted H steadily burns on the surface of the WD at the rate $\dot{M}_{\rm cr}$ and the unprocessed matter is lost from the system by the wind driven by the opacity arisen from the metal. However, for Pop III stars, such a scenario is unsuitable due to the accreted material's metal-free properties. \citet{meng2017} recently proposed the common envelope wind (CEW) model, which can be used to describe the SN Ia born at high redshift. In this model, if the mass transfer rate exceeds $\dot{M}_{\rm cr}$, the accumulated material onto the WD surface ultimately leads to the occurrence of the common envelope (CE). Due to the low density of the matter in the CE, the merger may not happen. The extended envelope material will produce the circumstellar medium with low velocities. \citet{meng2017} differentiated the CE into two regions, i.e., a rigid rotating inner region and a differentially rotating outer region. The outer CE will extract orbital angular momentum from the inner binary system, which affects the binary orbital evolution. 

In the CEW model, the accretion rate is set to be equal to the critical accretion rate once the CE forms. If the mass transfer rate is less than $\dot{M}_{\rm cr}$, the accumulation efficiency of WDs is computed with $\dot{M}_{\rm WD}=\eta_{\rm H}\eta_{\rm He}|\dot{M}_{\rm MT}|$, similar to that of the OTW model, where $\eta_{\rm H}$ {is the mass accumulation efficiency for H burning and is calculated by \citep{hachisu1999}} 
\begin{equation}
	\eta_{\rm{H}}=
	\begin{cases}
	\dot{M}_{\rm{cr}}/|\dot{M}_{\rm{MT}}|,\quad\quad |\dot{M}_{\rm{MT}}|>\dot{M}_{\rm{cr}} \\
	1,\qquad\qquad\quad \dot{M}_{\rm{cr}}\ge|\dot{M}_{\rm{d}}|\ge \frac 1 8\dot{M}_{\rm{cr}} \\
	0, \qquad\qquad\quad |\dot{M}_{\rm{d}}|<\frac 1 8 \dot{M}_{\rm{cr}},
	\end{cases}
	\label{eq:2}
\end{equation}
{and $\eta_{\rm He}$ is the mass accumulation efficiency for {He-shell flash} and its value is taken from \citet{kato2004}. The WD can accrete material efficiently if the mass transfer rate is in the range of ($1/8\dot{M}_{\rm cr},\;\dot{M}_{\rm cr}$), which is also known as stable-burning regime (see also \citealt{chenh2019}).} {The WD accretion processes also depend on the metallicity \citep{chenh2019}, here we assume that the accumulation efficiency for metal-free material is the same as that of solar metallicity (see also \citealt{meng2009}).}

The above physical processes are incorporated in the binary evolution code, and more than 1000 WD+MS binary sequences are simulated. The initial WD masses range from $0.9$ to $1.2M_\odot$ in steps of $0.1M_\odot$. {{For metallicity with $\lesssim 0.05Z_{\odot}$, the maximum masses of CO WDs generally have masses less than $1.1M_\odot$ \citep{mengx2008}. More massive WDs may ignite the center C and form ONe WDs. However, hybrid CO-Ne WDs may be produced due to the uncertainty of C-burning rate and the treatment of convective boundaries (e.g \citealt{chenm2014}). Hybrid WD could be as large as $1.3M_\odot$. The mass accretion of such type of WD may also lead to the thermonuclear runaway, resulting in the birth of SN Ia \citep{mengx2014}. Therefore, we choose the upper limit of $1.2M_\odot$ for the accretor to include the case of hybrid WD. In this work, we assume that the WD will explode as an SN Ia if the WD mass reaches $1.378M_\odot$} \citep{nomoto1984}. }The initial masses of MS stars range from $1.0M_\odot$ to $5.0M_\odot$ in steps of $0.2M_\odot$. The initial orbital periods range from $0.2$ to $120\;\rm d$ in steps of $\Delta \log P(\rm d) = 0.1$. We terminate the simulations if the mass transfer rate exceeds $10^{-3}M_\odot\;\rm yr^{-1}$, since dynamical unstable mass transfer is supposed to happen in such a high rate \citep{luw2023}. {All \texttt{MESA} input files are available at \dataset[doi:10.5281/zenodo.11218234]{https://doi.org/10.5281/zenodo.11218234}.}

\begin{figure*}
    \centering
    \includegraphics[width=0.85\textwidth]{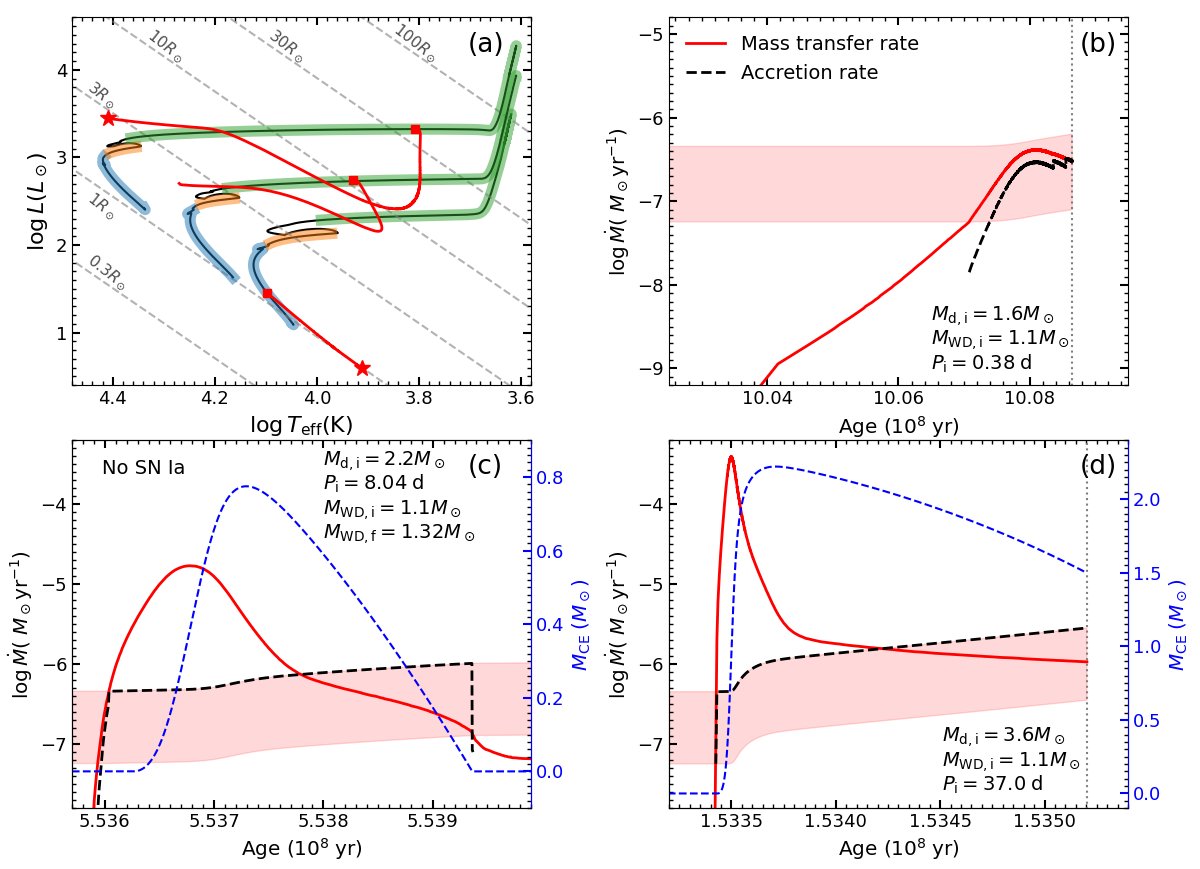}
    \caption{\textbf{Panel (a):} Single and Binary evolutionary tracks of Pop III stars. The solid black lines represent evolutionary tracks of single stars with masses of $1.6,\,2.4$, and $ 3.6M_\odot$ from the bottom to the top, respectively. Thick blue, yellow, and green lines represent the evolutionary stages of central H burning, H-shell burning, and He-shell burning, respectively. Three typical tracks of donor stars in binary systems are shown in red lines from bottom to top, respectively. The onsets of mass transfer are shown in red squares, and the onsets of SN explosions are shown in red stars. The grey dashed lines represent lines of constant radius. \textbf{Panel (b):} The accretion processes of WD during the mass transfer stages for the binary with initial parameters of $M_{\rm d,i}=1.6M_\odot,M_{\rm WD,i}=1.1M_\odot$, and $P_{\rm i}=0.38$ {d}. The accretion rate of WD and mass transfer rate are shown in red-solid lines and black-dashed lines. The stable burning regions are shown in the red-hatched region. \textbf{Panel (c):} Similar as panel (b), but for binary with initial parameters of $M_{\rm d,i}=2.2M_\odot,M_{\rm WD,i}=1.1M_\odot$, and $P_{\rm i}=8.4M_\odot$. The mass of the CE, $M_{\rm CE}$, is shown in a blue-dashed line. This binary system does not result in SN explosion eventually and the WD mass at the termination of mass transfer is $1.32M_\odot$. \textbf{Panel (d):} Similar as panel (c), but for binary with initial parameters of $M_{\rm d,i}=3.6M_\odot,M_{\rm WD,i}=1.1M_\odot,P_{\rm i}=37.0M_\odot$.}
    \label{fig:1}
\end{figure*}

\begin{figure*}
    \centering
    \includegraphics[width=0.85\textwidth]{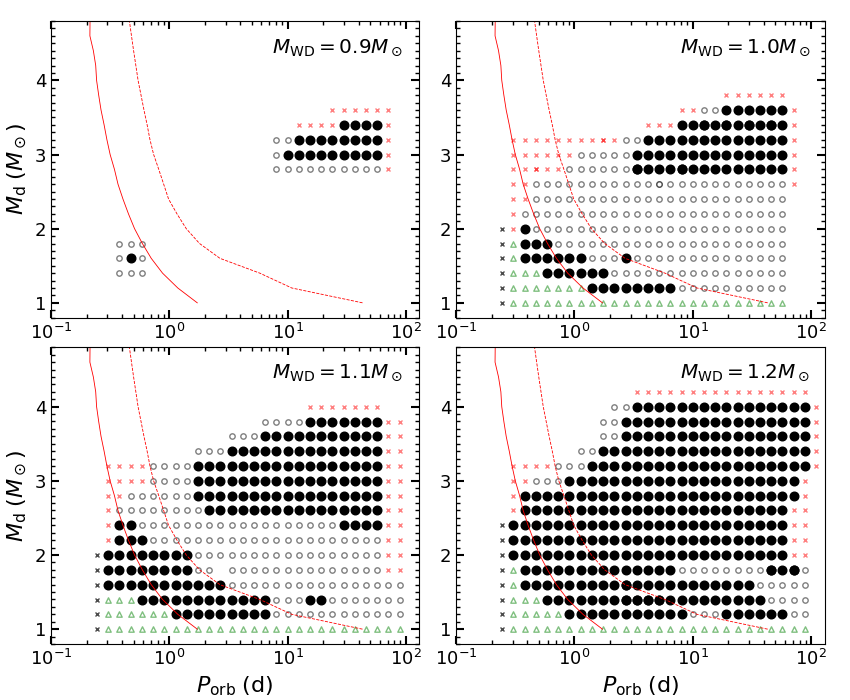}
    \caption{The initial parameter spaces for SNe Ia from Pop III stars. Black solid circles represent the progenitor binaries of SNe Ia. Green triangles indicate systems that experience nova explosions and the mass transfer rate is lower than the WD critical accretion rate. Open circles indicate that the mass transfer rate exceeds the WD critical accretion rate and the CE is formed, but the resulting CE is not massive enough that the WD cannot grow its mass to $1.378M_\odot$. The red crosses show the systems that experience dynamically unstable mass transfer and the black crosses in the lower left are for donors filling their Roche lobe initially. The red solid lines and dashed lines show the boundaries for models that start mass transfer at the termination of central H burning and at the onset of He core burning, respectively. The initial WD masses are shown in the upper right of each panel.} 
    \label{fig:2}
\end{figure*}

\subsubsection{Binary evolution results}

In the panel (a) of Figure \ref{fig:1}, we present three typical evolutionary tracks of Pop III stars from the onset of H burning to the late stage of He-shell burning in the Hertzsprung-Russell diagram, the stars have masses of $1.6,2.4,3.6M_\odot$ from bottom to top, respectively. The important stages, including the central H burning, central He burning, and the He shell burning, are shown in thick blue, yellow, and green lines, respectively. Due to the lack of metals, particularly the CNO elements, the central nuclear burning is only supported by the pp chains initially, and the star would have to be hotter and more compact to produce enough energy to counteract gravity. It is noted that during the center H burning, the CNO cycle reaction may happen once a little bit of carbon is produced via the 3-$\alpha$ reaction. The moment of the onset of the CNO cycle strongly depends on the stellar mass (see details in \citealt{marigo2001}). After the exhaustion of H in the center, the star expands due to the shell burning, similar to the Pop I star case. But the central He may be ignited during the shell burning stage for Pop III stars (the specific position also depends on the stellar mass; \citealt{marigo2001}). The central He burning leads to the radial contraction, and the star expands again at the late stage of central He burning. After the end of center He exhaustion, the subsequent expansion of the star is mainly supported by the combination He-shell (3-$\alpha$ reaction) and H-shell (CNO cycle) burning \citep{lawlor2023}. 


In a WD+MS binary system, the MS star expands and may fill its Roche lobe, leading to the mass transfer. Analogous to the Pop I binaries, we distinguish three cases of mass transfer, i.e., Case A, B, and C, corresponding to the onset of mass transfer at the core-H burning (thick blue line), shell-H burning (thick yellow line), and after core-He exhaustion (thick green line), respectively. Three typical binary evolutionary tracks are also presented, as shown in the red lines in Figure~\ref{fig:1}~(a). Successful explosions are found for Figures~\ref{fig:1}~(b) and (d). It is noted that the example with a $2.2M_\odot$ donor (Figure~\ref{fig:1}~(c)) will not produce SN Ia successfully before the termination of the accretion process. The onsets of mass transfer are shown in red squares, and the explosions of SNe Ia are shown in red stars. The details of the accretion processes for the WDs are shown in Figure~\ref{fig:1} (b-d) for the three examples with their initial parameters labeled in the Figure. The stable burning regions are shown in red-hatched areas. For the case of $M_{\rm d,i}=1.6M_\odot$, and $P_{\rm i}=0.38\;\rm d$ (Figure~\ref{fig:1}(b)), the initial thermal timescale mass transfer rate is in the stable burning region, the WD can accrete mass in a highly efficient way. The binary finally leads to the SN Ia explosion. With the increases of initial donor mass and the orbital period, the thermal timescale mass transfer exceeds the critical accretion rate, such as the case shown in panel (c), this results in the formation of a CE. In the CEW model, if the CE exists, the WD can accumulate masses with the critical mass transfer rate, which is always more efficient than the OTW model (see detailed discussions in \citealt{meng2017}). The WD then grows its mass at a rate of $\dot{M}_{\rm cr}$, meanwhile, the material is lost from the CE via the wind at a rate of several $10^{-6}M_\odot\;\rm yr^{-1}$ \citep{meng2017}. The WD mass increases to $1.32M_\odot$ as the CE disappears. In the subsequent evolution, the WD cannot accrete mass due to the strong shell flashes ($\dot{M}_{\rm MT}<1/8\dot{M}_{\rm cr}$). For the case in panel (d), where the binary has a more massive donor ($3.6M_\odot$), we see that the CE mass can grow to a maximum of $2.4M_\odot$. Finally, the WD mass increases to $1.378M_\odot$ when $M_{\rm CE}=1.5M_\odot$. The interaction between the massive CE and the SN explosion may leave footprints in the high-velocity features of SN Ia spectrum (see more details in \citealt{meng2017})

The corresponding parameter spaces leading to the SNe Ia for the different WD masses in the SD channel are shown in Figure~\ref{fig:2}. The parameter space is larger for more massive WD as expected, and the minimum WD mass of CO WDs leading to SNe Ia is 0.9 M$_\odot$, consistent with the deduction in \citet{meng2009}. {We also simulate the case of initial WD mass with  0.8 M$_\odot$, and no binary leads to the SN Ia explosion successfully.} For comparison, WD with mass as low as 0.65 M$_\odot$ can explode successfully as SN Ia for Pop I binaries (e.g., \citealt{meng2017}). The boundaries of the parameter space can be understood as follows. For low-mass donors and short orbital periods, the mass transfer rates are generally lower than $1/8$ $\dot{M}_{\rm cr}$, which prevents the WD accretion due to strong hydrogen shell flashes, as shown by the green triangles in Figure~\ref{fig:2}. The mass transfer rate is larger than the critical accretion rate for the cases shown by the open circles. However, most materials are lost via the CEW, similar to the example presented in Figure~\ref{fig:1}~(c), and the WDs cannot accrete enough masses to reach the Chandrasekhar mass limit. The red crosses represent the binaries experiencing dynamically unstable mass transfer, where the mass transfer rates reach $10^{-3}$ M$_\odot\;\rm yr^{-1}$ in the simulations. The open circles divide the parameter spaces for $M_{\rm WD}=0.9, \, 1.0$,  and 1.1 M$_\odot$ into two enclosed regions. In the lower part, the transferred masses can burn stably, similar to the case in Figure~\ref{fig:1}~(b). In the upper part, the high mass transfer rate leads to a massive CE and the WD can grow its mass through the CE, similar to the case in Figure~\ref{fig:1}~(d). For $M_{\rm WD}\ =\ 1.2\  {\rm M_\odot}$, the WD can explode as SN Ia with a small amount of the masses being accreted. Therefore, the open circles mainly appear at the lower right part, where the resulting CE masses are small. Such gaps in the parameter spaces are not found for Pop I stars \citep{meng2017}. Since binaries in the upper right part, i.e., stars with large masses and radii, are more likely to experience dynamically unstable mass transfer in the case of high metallicity \citep{geh2023}.

\subsection{Double Degenerate Channel}

In the double degenerate channel, the SNe Ia arise from the merger of two WDs with a combined mass exceeding the Chandrasekhar mass. A shred of important evidence favoring the DD scenario is that the predicted {delay time distributions (DTDs)} of SNe Ia in this channel could well reproduce that of the observations. Besides, some other observational clues, such as the absence of surviving companion stars in SN Ia remnants \citep{ruiz2018} and the null detection of the signatures of H/He spectral lines in the SN Ia nebular phase spectra \citep{leonard2007,maguire2016}, seem to support the DD scenario. In this {work}, we do not intend to perform the detailed population synthesis calculations for the DD channel, instead, we assume that SNe Ia from Pop III binaries share a similar DTD from Pop I stars. Such an assumption is justified because the delay times in the DD channel are mainly dominated by the merger timescale due to the {GW radiation}. The simulations in \citet{toonen2012} support this view and the effect of metallicity on SN Ia production is expected to be weak. We adopt the numerical results of the DTD of SNe Ia from the DD channel in the work of \citet{wangb2012}. We also include other power-law DTDs of SNe Ia, e.g., \citet{ruiter2009,toonen2012,claeys2014}, which give similar results.

\section{SN Ia Rate in Cosmological frame}
\label{sec:3}

\subsection{Cosmological parameters}
\label{sec:3.1}

Based on the parameter spaces of the SD channel and the DTD of the DD channel, one can estimate the SN rates. We calculate the {SN Ia rate} in the comoving frame with the semi-analytic method (\citealt{Santoliquido2020,Santoliquido2021,santoliquido2023}), which is given by 
\begin{eqnarray}
  \mathscr{R}(z) = \int_{z_{\rm max}}^{z}\psi(z')\left[\frac{\nu(z',z)}{{\rm d}t(z')}\right]\frac{{\rm d}t(z')}{{\rm d}z'}{\rm d}z',
  \label{eq:3}
\end{eqnarray}
where $\psi(z')$ is the star formation rate density at $z'$, {$\nu(z',z)$ is the number of SNe Ia per M$_\odot$ with the birth of progenitor at redshift $z'$ and the birth of SN Ia at $z$. The detailed methods to obtain $\nu$ are given in Section~\ref{sec:3.2}}. The standard $\Lambda$CDM cosmological model is adopted (e.g., \citealt{peebles1993,bullock2017}), in which the Universe is composed of the cold dark matter, baryons, and the dark energy. It gives ${\rm d}t(z')/{\rm d}z' = H_{0}^{-1}(1+z')^{-1}[(1+z')^3\Omega_{\rm M}+\Omega_{\Lambda}]^{-1/2}$, where $H_0=67.4\;\rm km/s\;Mpc^{-1}$ is the Hubble parameter, $\Omega_{\rm M}=0.3153$ and $\Omega_{\Lambda}=0.6847$ are the matter and energy density with values taken from \citet{Planck2020}. 
The CSFH is one of the most important parameters determining the properties for an assigned binary population \citep{santoliquido2023}. There are many simulations concerning the CSFH models of Pop III stars (e.g., \citealt{jaacks2019,liub2020,skinner2020,hartwig2022}). {In this work, we mainly consider two representative models of CSFH.} One with the relatively lower star formation rate is taken from \citet{liub2020} (hereafter LB20 model), i.e.,  
\begin{eqnarray}
  \psi(z) = \frac{a(1+z)^b}{1+[(1+z)/c]^d}\;\;[M_\odot\rm yr^{-1}Mpc^{-3}],
  \label{eq:4}
\end{eqnarray}
where best-fit parameters $a=756.7M_\odot\rm yr^{-1}Mpc^{-3},\;b=-5.92,\;c=12.83$ and $d=-8.55$. The second is taken from \citet{sarmento2019}, who performed large-scale cosmological simulations to get star formation rate as a function of $z$ (hereafter {S19} model). S19 model predicts a peak of CSFH at $z\sim 8.5$ and the simulations was run down to $z=7$ with a relatively higher CSFH compared to the LB20 model. {The CSFHs from several other studies are also introduced in Appendix~\ref{app:B}, where the influence of CSFHs on the results is briefly addressed.}

\subsection{Initial Distributions for Binary Parameters}
\label{sec:3.2}

The calculations of the SN Ia rate are divided into two steps, corresponding to the SD channel and DD channel. For the SD channel, we have obtained the grids that produce SNe Ia as shown in Figure \ref{fig:2}. {Then we need to transform the parameter spaces to the initial parameter ranges of the primordial binary systems. We first introduce the method to obtain the contribution to SN Ia rates for the relevant distributions of the primary mass, binary mass ratio, and orbital separation.} 

{If a particular primordial binary system $j$ with initial binary parameters of primary and secondary mass $M_{1,j}$ and $M_{2,j}$, and orbital separation $A_j$ evolves to an SN Ia successfully (the binary system $j$ evolves to a grid point in the regions of the black solid circles in Figure~\ref{fig:2}), then its contribution to the production of the SNe Ia is given by \citep{Hurley02,Shaoy2021}}
\begin{align}\label{eq:5}
  \delta\nu_{j} &= X\left(\frac{2f_{\rm b}}{1+f_{\rm b}}\right)\left(\frac{1}{M_{*}}\right)\chi(\ln M_{{\rm 1},j})\varphi(q_{j})\Psi(\ln A_{j})\\
  &\times\delta \ln M_{\rm 1} \cdot \delta q \cdot \delta \ln A \;\;[{\rm numbers \; per}\;M_\odot],\nonumber
\end{align}
{where $X$ is factor of order unity to account for the uncertainties of SN Ia models in comparison with the observations, its value is given in Section~\ref{sec:4.1}, $f_{\rm b}$ is the binary fraction among all the stars. In our standard models, we assume that all stars are in binaries, i.e., $f_{\rm b}=1$, which gives an upper limit of the SN rates. $M_*$ is the mean mass of the binary systems, $q_{j}$ = $M_{2,j}/M_{1,j}$ is the mass ratio, }
{and $\chi(\ln M_{1,j})$, $\varphi(q_{j})$ and $\Psi(\ln A_{j})$ are normalization functions to weight the contribution of the specific binaries with initial parameters of $\ln M_{1,j}$, $q_{j}$ and $\ln A_{j}$, respectively. Thus }
\begin{eqnarray}
    \chi(\ln M_{\rm 1,j}) = M_{1,j}\xi(M_{\rm 1,j}), 
    \label{eq:6}
\end{eqnarray}
{where $\xi(M)$ is the initial mass function (IMF).}
{We assume a uniform distribution of the mass ratio \citep{mazeh1992}, i.e., }
\begin{eqnarray}
    \varphi(q_{j}) = 1, \;0<q_{j}\leq 1.
    \label{eq:7}
\end{eqnarray}
{The initial  distribution of the separation is assumed to be \citep{popova1982,abt1983}}
\begin{eqnarray}
    \Psi(\ln A_{j}) = k, \; 10<A_{j}<10^4 R_\odot,
    \label{eq:8}
\end{eqnarray}
{where $k=0.14476$ is a normalized coefficient. 
According to equation~(\ref{eq:6}-\ref{eq:7}), the mean mass of all of the binaries is}
\begin{eqnarray}
    M_* = \frac 3 2 \int_{M_{\rm low}}^{M_{\rm up}} M\xi(M){\rm d}M,
    \label{eq:9}
\end{eqnarray}
{where $M_{\rm low}$ and $M_{\rm up}$ are the lower and upper limit of stellar masses for a given IMF. }

Very little is known about the IMF for Pop III stars. Early numerical simulations favored a top-heavy IMF with most stars more massive than $\sim 100M_\odot$ (e.g., \citealt{omukai1998,abel2002,bromm2002}). However, many recent works on the formation of Pop III stars suggest that fragmentation could lead to a large number of low- and intermediate-mass stars ($\lesssim 10$ M$_\odot$; e.g., \citealt{jaura2022,prole2022}). Since SNe Ia are produced by these low- and intermediate-mass stars, as shown in Figure \ref{fig:2}, it can be foreseen that the choice of the IMF can significantly affect the prediction of SN Ia rate. We consider the following two forms of the IMF for Pop III stars: 

\begin{itemize}
\item[1.] {Bottom-heavy function the same as Pop I IMF (e.g., \citealt{kroupa1993}), i.e., 
\begin{eqnarray}
    \xi(M)=
    \begin{cases}
    0,&M\leq0.1M_\odot \\
    a_1M^{-1.3}, &0.1<M\leq0.5M_\odot \\
    a_2M^{-2.3}, &0.5<M\leq1M_\odot \\
    a_2M^{-2.7}, &1<M\leq100M_\odot
    \end{cases}
    \label{eq:10}
\end{eqnarray}
where $a_1 = 0.29056$ and $a_2=0.15571$. Insert equation~(\ref{eq:10}) into equation~(\ref{eq:9}), we get $M_*=0.75M_\odot$. }

\item[2.] {Top-heavy function for Pop III stars according to several numerical simulations} (e.g., \citealt{stacy2013,susa2014,stacy2016,wollenberg2020,sharda2020,jaura2022,prole2022}), i.e., 
\begin{eqnarray}
    \xi(M) = a_3M^{-1},
    \label{eq:11}
\end{eqnarray}
{where the normalized factor $a_{3}$ depends on the mass range of Pop III stars.} 
For a top-heavy function, massive stars dominate the stellar masses. Therefore, the upper limit of the star mass ($M_{\rm up}$) significantly affects the numbers of primordial binaries with masses in the range of $\sim 1-10M_\odot$ (typical mass for the WD progenitor). {We choose $M_{\rm low}=0.1M_\odot$ and consider three cases with $M_{\rm up}=10,\,50,\,$ and  500 M$_\odot$. The corresponding values of $a_3$ for these three cases are $0.21715$, $0.16091$, and $0.11741$, and the corresponding values of $M_{*}$ are $3.22M_\odot$, $12.04M_\odot$, and $88.04M_\odot$, respectively. The total stellar masses in the range of $1-10$ M$_\odot$ of these three cases are $90\%,\, 18\%,$ and  $2\%$, respectively.}

\end{itemize}

{The methods to transform the parameter spaces of Figure~\ref{fig:2} to the initial parameter ranges of the primordial binary systems are given as follows. For each grid point that produces SNe Ia successfully (black solid circles in Figure~\ref{fig:2}), we first find the typical mass of the initial primary star (WD progenitor). Here the WD masses of 0.9, 1.0, 1.1, and 1.2 M$_\odot$ are assumed to correspond to the four mass intervals $0.85-0.95$ M$_\odot$, 0.95-1.05 M$_\odot$, 1.05-1.15 M$_\odot$, and $1.15-1.25$ M$_\odot$, respectively. The weights of the corresponding mass intervals is 
\begin{eqnarray}
    \Delta M_{1,j} = \int_{M_{\rm A}}^{M_{\rm B}}\xi(M){\rm d}M,
    \label{eq:12}
\end{eqnarray}
where $M_{\rm A}$ and $M_{\rm B}$ are the typical progenitor (primary) masses of WDs at each end of the WD mass intervals. The methods to get the corresponding masses of WD progenitors are presented in Appendix~\ref{app:A}. 
}

{The weights of the corresponding mass ratio intervals is \citep{hachisu1999,chenx2011}}
\begin{eqnarray}
    \Delta q_{j} = \frac{M'_{\rm u}}{M_{\rm A}}-\frac{M'_{\rm l}}{M_{\rm B}},
    \label{eq:13}
\end{eqnarray}
{where $M'_{\rm u}$ and $M'_{\rm l}$ are the upper and lower limits of the initial donor (secondary) masses that produced SNe Ia successfully in Figure~\ref{fig:2}. In our simulations, the adopted mass interval of $M_{\rm d}$ is $0.2M_\odot$. Therefore, for a grid point $j$ with a donor mass of $M_{{\rm d},j}$, $M'_{\rm u}$ and $M'_{\rm l}$ equal to $M_{{\rm d},j}+0.1M_\odot$ and $M_{{\rm d},j}-0.1M_\odot$, respectively, and $M_{{\rm d},j}$ can be identified directly from Figure \ref{fig:2} (see also \citealt{hachisu1999,chenx2011} for more details).}

{The weights of the corresponding separation intervals is}
\begin{eqnarray}
   k\Delta \ln A_{j} = k\Delta \ln a_{j} = \frac 2 3 k \Delta \ln P_{\rm orb},
    \label{eq:14}
\end{eqnarray}
{where $k$ is the normalization coefficient in equation~(\ref{eq:8}), $A_{j}$ is the initial separation of the primordial binary, $a_{j}$ is the initial separation of the WD+MS binary, $P_{\rm orb}$ is the orbital period taken from the SN Ia region in Figure~\ref{fig:2}, the factor of $2/3$ comes from the conversion between the period and the separation. Equation~(\ref{eq:14}) implies that the difference between $\Delta \ln A_{j}$ and $\Delta \ln a_{j}$ is unchanged by the evolution up to the point of Figure~\ref{fig:2}, i.e., the probability frequency for $\Delta \ln A_{j}$ is the same as for $\Delta \ln a_{j}$. The physical motivation of the above assumption is that WD+MS binaries in the parameter spaces of Figure~\ref{fig:2} generally have gone through one CE (here the CE means the standard CE phase, i.e. an evolved star engulf its companion, which is different from the CE in CEW model discussed above) ejection process \citep{willems2004}, and the orbital separation has shrunk by a contraction factor after the CE ejection (see more detailed discussions about this issue in \citealt{hachisu1999}).}

{In combination with equations~(\ref{eq:6}-\ref{eq:14}), equation~(\ref{eq:5}) now becomes }
\begin{eqnarray}
    \delta \nu_{j} = X\cdot\left(\frac 1 {M_*}\right)\cdot k \cdot \Delta \ln M_{ 1,j}\Delta q_{j}\Delta \ln a_{j},
    \label{eq:15}
\end{eqnarray}
{the total contribution to SN Ia rate according to the grid (black solid circles) in Figure~\ref{fig:2} is }
\begin{eqnarray}
    \nu = \sum_{j} \delta \nu_{j}.
    \label{eq:16}
\end{eqnarray}

{In this work, we obtain the SN Ia rate based on a semi-analytic way. Therefore, several complicated distributions of initial binary parameters are hard to take into account. We plan to perform detailed binary population synthesis in a future study.}

\subsection{Routes to SN Ia rate}
\label{sec:3.3}

The {SN Ia rate} from the SD channel is then calculated according to equation~(\ref{eq:15}), where the parameter ranges that produce SNe Ia can be directly read from Figure \ref{fig:2}. When inserting the value of $\nu$ into equation (\ref{eq:3}), we also need to know the delay time, i.e., the time elapsed from the primordial binary to the birth of SN Ia. The delay time is approximately divided into two parts: the birth time of the massive WD and the evolutionary timescale of the secondary ({donor star of WD+MS binary}) until the explosion of SNe. The WDs with masses of $0.9-1.2$ M$_\odot$ ({corresponding to the extending parameter range of $0.85-1.25$ M$_\odot$}) have progenitor masses in our grid about $\sim\, 3.7-7.3$ M$_\odot$, the corresponding evolutionary timescale is about $3.86\times 10^7-1.17\times 10^8$ yr ({see Appendix~\ref{app:A}}). The secondary stars that produce SNe Ia have masses in the range of $1.2-4.0$ M$_\odot$, corresponding to a timescale of $\sim 1.2\times 10^8-3.4\times10^9$ yr. For each grid point in Figure \ref{fig:2}, we sum the two parts of time scales to get the delay time and then obtain the SNe Ia rate as a function of $z$ according to equation (\ref{eq:3}).

\begin{figure}
    \centering
    \includegraphics[width=\columnwidth]{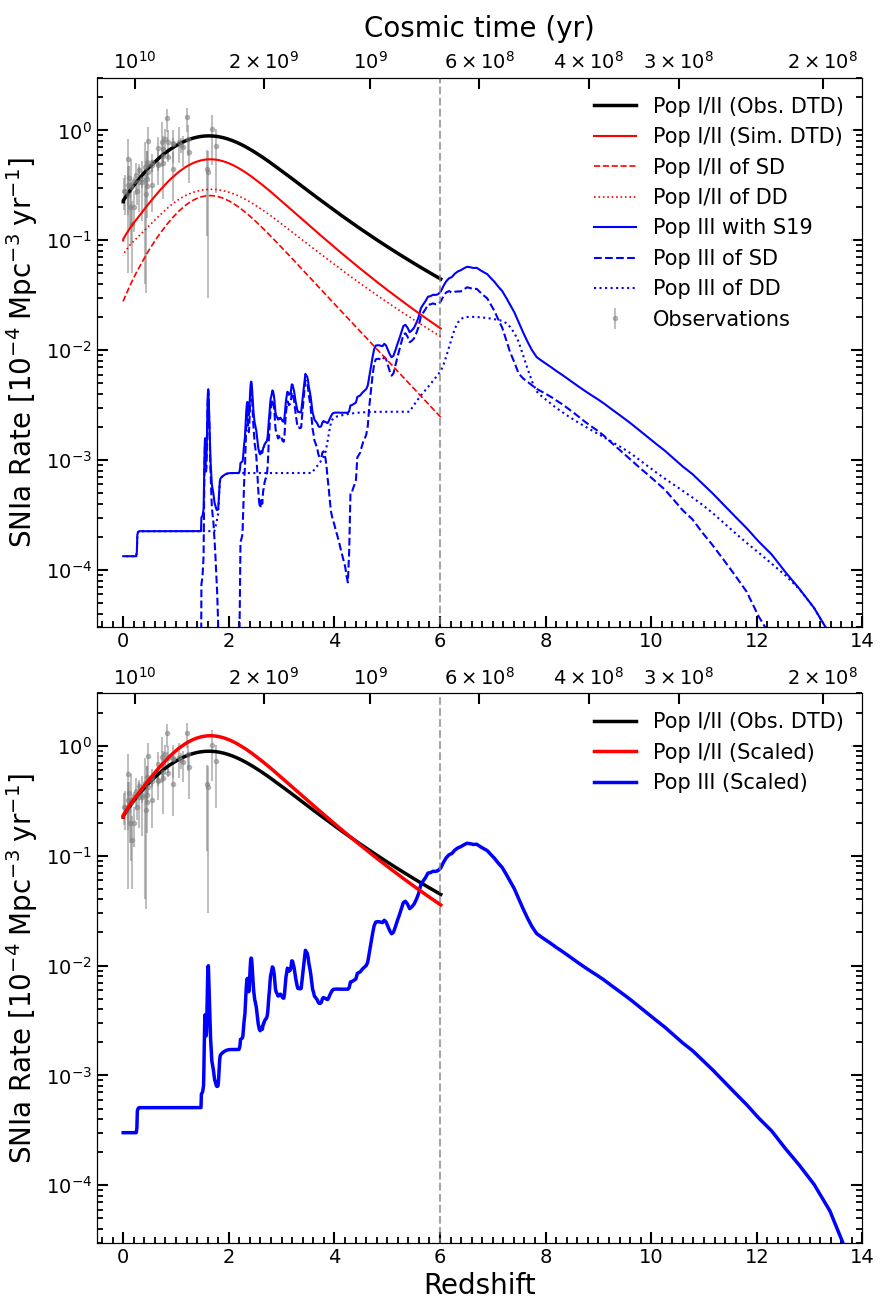}
    \caption{\textbf{Upper panel:} The birthrate of SNe Ia as a function of redshift. The total birthrate of SNe Ia from Pop III stars of {S19} model is shown in blue solid line, and the contributions of SD and DD channels are shown in blue dashed and dotted lines, respectively. The observed SN Ia rates at $z<2$ are shown in grey dots with error bars (collected by \citealt{strolger2020}). The black solid line is the best-fit curve obtained with the convolution of Pop I/II CSFH and a power-low DTD from \citet{maoz2017}. {We make a cutoff for Pop I/II SN Ia rates at $z=6$.} The simulated SN Ia rate with theoretical DTDs of Pop I stars are shown in the red solid line, where the contributions of SD and DD channels are shown in red dashed and dotted lines, respectively. The observed rates are about twice as high as the rates from theoretical simulations mainly because the DTD in the observations is generally larger than the theoretical simulations. The corresponding cosmic time is shown in the upper axis. See more details in the text. \textbf{Lower panel:} {The scaled SN Ia rate as a function of redshift. To match the observations, we artificially scale up the SN Ia rates by a factor ($2.2$), so that the fitted and theoretical curves of Pop I/II SN Ia rates have the same value at $z=0$}.}
    \label{fig:3}
\end{figure}

In the DD channel, we directly insert the adopted DTD of SNe Ia into equation (\ref{eq:3}). Since we do not perform the binary population synthesis for SNe Ia from Pop III binaries, the DTD of SNe Ia from the DD channel is adopted with the same profile as that for Pop I binaries, i.e., the numerical result of \citet{wangb2012}. {However, the effect of alternative initial binary parameter distributions different from the ones considered in this work is hard to quantify due to the lack of detailed binary population synthesis calculations. Based on the results of the SD channel, we found the SN Ia rate scales approximately as a weak function of $z$ for different assumptions of the IMFs (as shown below). We thus apply a constant scaling factor in calculating the SN Ia rate in the DD channel, where the scaling factor is given by dividing the mean value of SN Ia rate for a specific IMF by that for the bottom-heavy IMF in the SD channel. }

The total SN Ia rate is the sum of the SD and DD channel contributions. 

\section{Results}
\label{sec:4}

\subsection{SNe Ia from different channels}
\label{sec:4.1}

{In the upper panel of Figure~\ref{fig:3}}, we present an example illustrating the methods to get the SN Ia rate. The results of SNe Ia from Pop I/II stars are also shown for comparison. The black line in the Figure is the best-fit curve obtained by convolving the CSFH of Pop I/II stars taken from \citet{madau2017} with a power low DTD {(index of $-1.07$; \citealt{maoz2017}) according to equation~(\ref{eq:3}).} The red dashed line and dotted line are for SNe Ia from SD channel and DD channel of Pop I stars, respectively, where the DTDs are taken from the binary population synthesis results from \citet{meng2017} and \citet{wangb2012}, respectively. {
With the assumed DTD of SNe Ia, the SN Ia rate as a function of $z$ is directly obtained from equation (\ref{eq:3}) without using equation (\ref{eq:5}).} The red solid line is the total SN Ia rate in the simulations (the sum of the dashed and dotted lines). It is noted that the rate from observational data (black solid line) is about two times higher than the theoretical simulations (red solid line). The reason is that the observed {DTDs} of SNe Ia are generally higher than the predicted {values} in theoretical simulations, as shown in many previous works (e.g., \citealt{toonen2012,claeys2014,liud2018,lizw2023}). {To match the observations, we artificially scale up the theoretical Pop I/II SN Ia rate by a factor of $2.2$, i.e., $X=2.2$ in equation~(\ref{eq:5}), so that the fitted and theoretical curves have the same value at $z=0$, as shown in the lower panel of Figure~\ref{fig:3}.}

The SNe Ia produced from Pop III stars are also shown in blue lines in the upper panel of Figure~\ref{fig:3}, where the {S19} model of the CSFH and top-heavy IMF {with $M_{\rm up}=10M_\odot$} are adopted. The dashed and dotted lines are for the SD and DD channels, respectively. The total SN Ia rate is shown in the blue solid line. For {S19} model, the star formation rate increases since the birth of the first star, {peaks around $z\sim 8.5$, and then} runs down to $z=7$, leading to a noticeable peak SNe Ia rates at $z \sim 6-7$. The multiple plateaus in the DD channel are caused by the large bin size of the delay time adopted. {The rapid fluctuations for the SD channel at $z\sim 1-4$ are due to the large mass interval of $M_{\rm d}$ in the grids of Figure~\ref{fig:2}. The sparsity of $M_{\rm d}$ then leads to the discontinuity of the delay time (the delay time is mainly determined by the lifetime of the donor star in the WD+MS binary, see Section~\ref{sec:3.3} for more details). In the S19 model, there is no Pop III star formation at $z\lesssim 7$. Therefore, discontinuity of the delay time would directly affect the SN Ia rate according to Equation~(\ref{eq:3})}. We also note that the SN Ia rate of the SD channel is {slightly} higher than that of the DD channel for Pop III binaries with $z\gtrsim 2$, while the case of Pop I/II is the opposite. {The main reason is that the parameter spaces that produce SNe Ia of Pop III stars are approximately $1-2$ times larger than that of Pop I stars when comparing our grids in Figure~\ref{fig:2} with Figure 7 of \citet{meng2017}. Besides, our grid spaces in Figure~\ref{fig:2} are not dense enough (especially for the WD mass interval), which may introduce a moderate overestimate of the SN Ia rate \citep{wangb2012}.}

{In the lower panel of Figure~\ref{fig:3}, we scale up the total Pop III SN Ia rate by the same factor as that used in matching the observed and theoretical SN Ia rates for Pop I/II stars. In the rest of the paper, all Pop III SN Ia rates are scaled by this factor due to the theoretical incompleteness of SN Ia formation models. The predicted SN Ia rates with other CSFHs can be found in Appendix~\ref{app:B}. }

\subsection{SN Ia rates in different models}

\begin{figure*}
    \centering
    \includegraphics[width=0.85\textwidth]{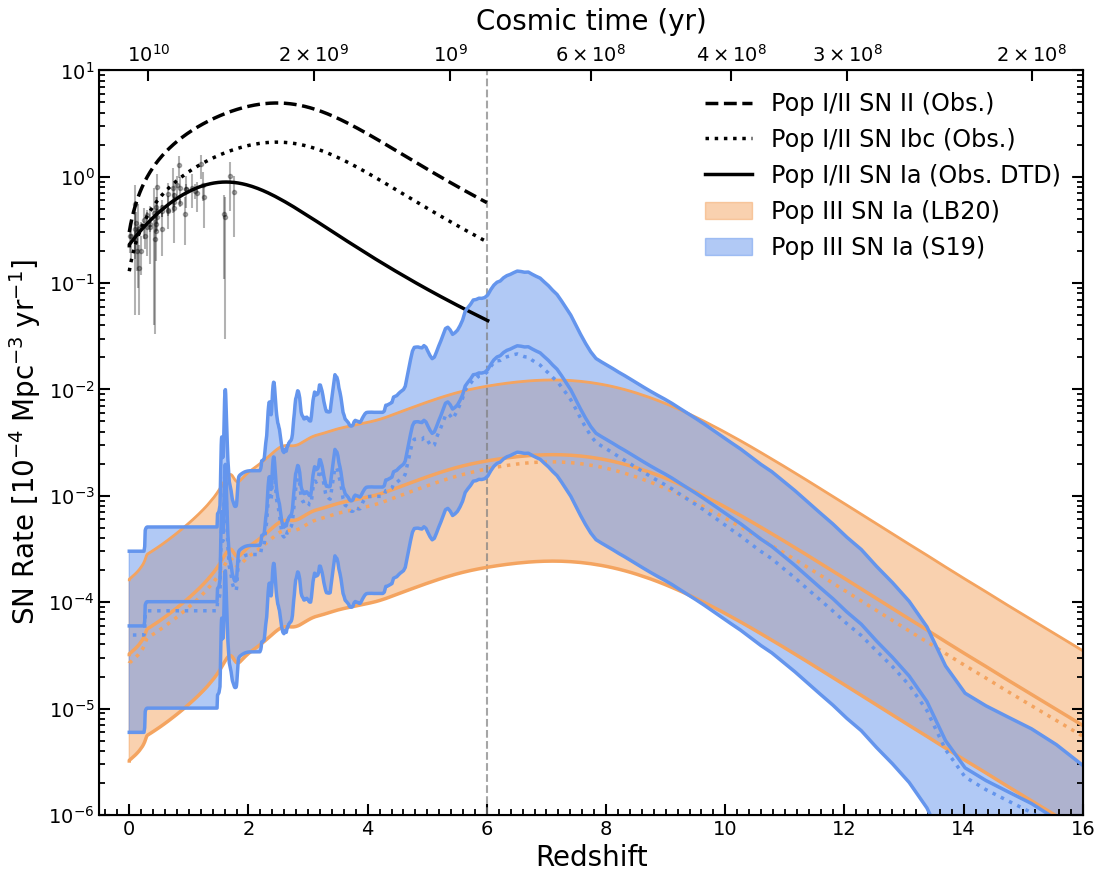}
    \caption{{Rates for a variety of SN spectral Types and models as a function of redshift.} The SN II and SN Ibc rates from Pop I/II stars are shown as the black dashed line and black dotted line, respectively, where the values at high redshift ($z\gtrsim 2$) are extrapolations based on the Pop I/II CSFH (taken from \citealt{wangl2017}). The black solid line is for Pop I/II SNe Ia based on the observed SN Ia rates {truncated at $z=6$.} The blue and yellow hatched regions are for Pop III stars with CSFH of {S19} and LB20 models, respectively. The solid and dotted lines in the hatched regions are obtained by using bottom-heavy and top-heavy IMFs, where the three solid lines from top to bottom are for the top-heavy IMF with $M_{\rm up}=10,\;50\;$, and 500 M$_\odot$, respectively. The corresponding cosmic time is shown in the upper axis. The contributions from individual formation channels are not presented for clarity.}
    \label{fig:4}
\end{figure*}

The prediction of {rates for a variety of SNe} as a function of redshift for a variety of SNe and models are shown in Figure \ref{fig:4}. For clarity, the contributions of individual formation channels are not presented. The blue and yellow hatched regions are for Pop III stars with CSFH of {S19} and LB20 models, respectively. The solid and dotted lines in the hatched regions are obtained by using {top-heavy and bottom-heavy} IMFs, respectively, and the three solid lines from top to bottom are for the top-heavy IMF with $M_{\rm up}$ $=10$, 50, and 500 M$_\odot$, respectively. The theoretical predictions of SN II rate and SN Ibc rate based on the volume-limited sample \citep{bazin2009,shivvers2017} are also shown for comparison \citep{wangl2017}. 

For both CSFH models, the SN Ia rates peak at $z$ around $\sim 6.5-8$ and then rapidly decreases in the nearby Universe. The trend is consistent with the Pop III star formation history, where very few Pop III stars are born at low redshift (no Pop III stars form at {$z\lesssim7$ in S19 model}). The peaks arise from the MS stars with mass $\gtrsim 2.8M_\odot$, which have delay timescale in the range of $1.2\times10^{8}-3.5\times10^{8}\;\rm yr$. The S19 model generally predicts more SNe Ia than the LB20 model at $z\sim 3-8$ due to the large star formation rate. We note that the SN Ia rate changes by about two orders of magnitude with different choices of IMFs. For the bottom-heavy IMF, most masses are in low-mass stars ($\lesssim2M_\odot$). Since the SN Ia progenitor (WD) masses are mainly in the range of $\sim 3.7-7.3M_\odot$, the bottom-heavy IMF would lead to a relatively lower SN Ia rate, as shown in the blue and yellow dotted lines. For the top-heavy IMF, though the number of massive stars significantly increases, most of the masses are in high-mass stars. Therefore, the choice of $M_{\rm up}$ determines how many of the total masses are in the range of $\sim 3.7-7.3M_\odot$. In the pessimistic case of $M_{\rm up}=500$ M$_\odot$, the SN Ia rate is even lower than that of the bottom-heavy IMF, as shown in the bottom blue and yellow solid lines. In the optimistic case of $M_{\rm up}=10$ M$_\odot$, i.e., $90\%$ of the masses are in the range of $1-10$ M$_\odot$, the SN Ia rate of Pop III stars {is comparable to or higher than} the SN Ia rate of Pop I/II stars at redshifts $z\sim 5-6$. 

The results are extremely sensitive to the IMF. In other words, observations of SN Ia at the redshift above $z\sim 6$ may provide stringent tests on the models of formation of Pop III stars at high redshifts. 

{The binary fraction of Pop III stars is also an important parameter affecting the SN Ia rate estimation. In this work, we assume that all stars are in binaries, which gives an upper limit of SN Ia rate for Pop III stars. If a lower binary fraction ($f_{\rm b}$) is adopted, the predicted rates would be reduced by a factor of $(1+f_{\rm b})/2f_{\rm b}$, where $f_{\rm b}$ is assumed to be a constant \citep{Shaoy2021}. For example, if $f_{\rm b}=0.5$ and $0.2$, the predicted SN Ia rates would reduce by $1.5$ and $3$ times, respectively. The binary fraction for Pop III stars is hard to constrain due to the lack of observations \citep{greif2011,greif2012}. As a comparison, more than $50\%$ Pop I stars are in binary systems (e.g., \citealt{sana2012,moe2017}). For OB-type stars, the binary fraction is as high as $80\%$ \citep{guoy2022}. In this case, our predictions on the SN Ia rates from Pop III stars are not significantly affected. 
}

\begin{figure}
    \centering
    \includegraphics[width=\columnwidth]{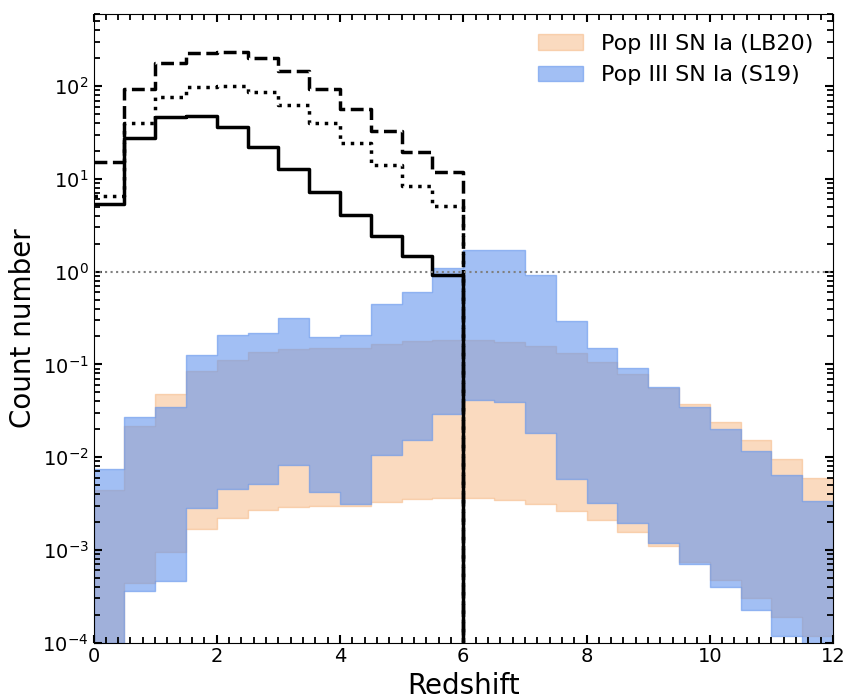}
    \includegraphics[width=\columnwidth]{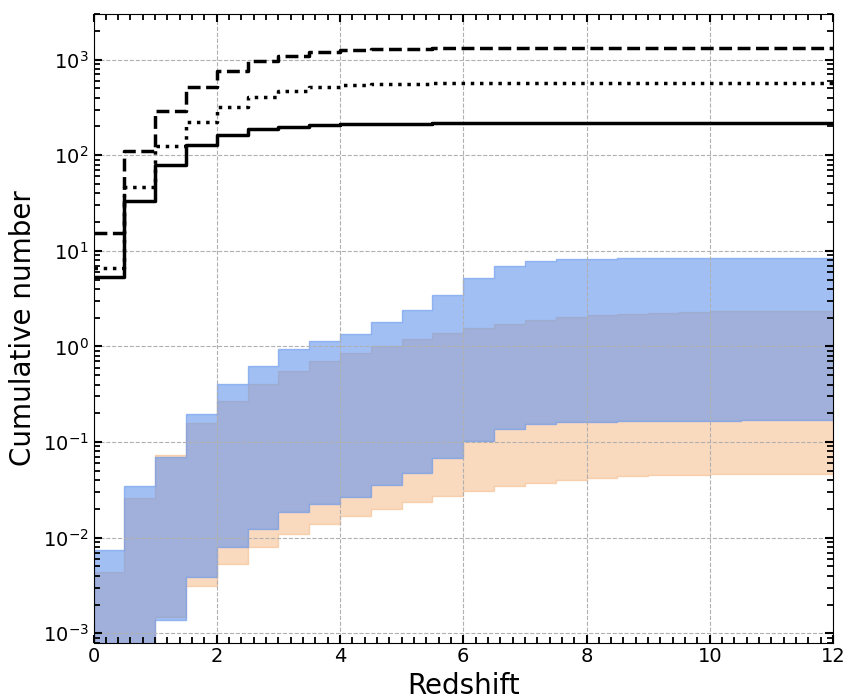}
    \caption{\textbf{Upper panel:} The expected number for JWST adopting a survey area of $300\;\rm arcmin^2$ and survey time of $3\;\rm yr$. The numbers of SNe II, SNe Ibc, Pop I/II SNe Ia with observational DTD are shown in black dashed, black dotted, and black solid lines, respectively. The Pop III SNe Ia of {S19} and LB20 models are shown in blue and orange hatched regions, where the upper boundaries are for the optimistic model of top-heavy IMF, and the lower boundaries are for the pessimistic model of top-heavy IMF with $M_{\rm up}=500$ M$_\odot$. \textbf{Lower panel:} Similar to the upper panel, but for the cumulative numbers. }
    \label{fig:5}
\end{figure}

\subsection{Predictions for a Wide Field Survey with the JWST}
Time-Domain survey with the JWST can discover transients in the distant universe. As an example, the FLARE project with JWST could map a field of view around $300\ \rm arcmin^2\,(0.083\,\rm deg^2)$ to $\sim 27.5$ AB-magnitude {using the Near Infrared Camera (NIRCam) of JWST \citep{wangl2017}, with a cadence of $\sim$ 90 days.} 
The expected number of SNe for such a survey can be calculated via 
\begin{eqnarray}
  N_{\rm SN} = \Omega T_{\rm obs}\int_{\rm z}^{\rm z+\Delta z}\frac{\mathscr{R}(z)}{1+z}\frac{{\rm d}V}{{\rm d}z}{\rm d}z,
  \label{eq:17}
\end{eqnarray}
where $\Omega\equiv0.083\,\rm deg^2$ is the survey area of FLARE project and $T_{\rm obs}$ $\sim 3$ years is the total survey time, and ${\rm d}V/{\rm d}z$ is the comoving volume \citep{regos2019}. 
{As in \citet{regos2019}, we adopted the model spectral energy distributions (SEDs) of SNe Ia close to the maximum light to predict the expected fluxes and the epoch of maximum light for each SN Ia is assumed to be distributed uniformly within the $3$ yr survey window. In the optimistic case, an SN Ia can be identified when at least one JWST NIRCam bandpass is detected at a given epoch. }
{The FLARE proposal has a limiting magnitude of $\sim$ 27.5 for most of the NIRCam filters which can detect SNe Ia at z $\lesssim 5$. However, SNe Ia out to z $\gtrsim 8$ can be detected if the depth of the survey is increased to a limiting magnitude of $\sim$ 28-28.5 thanks for favorable $K$-corrections in the F444W-band \citep[see Figure~5 of][]{wangl2017}. The F444W-band captures the rest frame $R$-band for the SNe Ia at $z > 6$ and allows for relatively reliable determinations of the SNe based on photometric light curves. The null detection in the F200W bands would eliminate SNe at lower redshifts \citep[see Figure~5 of][]{wangl2017}. }

{We thus assume that the JWST is sensitive enough to detect SNe Ia out to z well above 8.}
The count and cumulative numbers for different types of SNe are summarized in the upper and lower panels of Figure \ref{fig:5}, where the bin of $\Delta z = 0.5$. The upper boundaries for the hatched regions are for the expected numbers of SNe Ia for Pop III stars with {top-heavy IMF of $M_{\rm up}=10$ M$_\odot$}, which gives an upper limit for the prediction, and the lower boundaries are for the pessimistic model with the top-heavy IMF having $M_{\rm up}=500$ M$_\odot$. SNe II and SNe Ibc dominate all redshift ranges by several orders of magnitudes due to the large parameter ranges of their progenitors \citep{hopkins2006}. {In the upper panel, the expected numbers of SNe Ia for Pop III stars at $z\sim 5.5-7.5$ may be higher than $1$ in the S19 model. 
The bulge of SN Ia numbers at $z\sim 5.5-7.5$ is directly correlated with the peak of S19 CSFH at $z\sim 8.5$ (see Figure~\ref{fig:B1} in Appendix~\ref{app:B}). The detection of SNe Ia at high redshifts would place direct constraints on the CSFH. The count numbers of Pop III SNe Ia in a redshift bin at low redshift are generally less than $1$.} 

{In the lower panel, the upper limit of the cumulative number of SNe Ia for Pop III stars is about $1$ at $z\sim 4$,  $2$ at $z\sim 5$, and 8 for $z \gtrsim 8$. We may expect at most $\sim 2$ SNe Ia from Pop III stars in the FLARE project with limiting magnitudes of 27.5, but the number may quadruple for limiting magnitudes of 28.5.} Besides, one could expect more SNe Ia from Pop III stars with an extensive survey area and an extended survey time. Furthermore, the recent observations found an unexpected overdensity of galaxies at $z \sim 10$ \citep{castellano2023,yanh2023}, suggesting that the star formation rate at the high redshift could be significantly higher than previously believed. Therefore, there may be significantly more SNe Ia than our predictions if more Pop III stars are born at high redshift. 

It is worth noting that at low redshift ($z \lesssim 2.5$), the number of SNe Ia produced from the Pop I/II stars is higher than that of Pop III by about two orders of magnitude ($\sim 400$), which means that for a complete survey out to $z \sim 2.5$, there may be one Pop III SN Ia out of $\sim 400$ SN Ia events found at low redshifts. {A large number of SNe Ia, well above $1000$, have been discovered in recent years in the redshift range below 2.5 \citep[e.g.,][]{scolnic2018,des2024}. None of these SNe Ia is likely to be from Pop III as most of these discoveries are for SNe at $z$ well below 1. Even the Rubin Observatory’s LSST \citep{lsst2009} is likely to record a few SNe Ia from Pop III stars with its tremendous discovery capacity. 
However, it is a formidable task to identify Pop III SNe Ia from the enormous samples. A possible way is to use the heavy elements of the explosive nucleosynthesis to distinguish Pop III SN Ia \citep{leung2018,leung2020}. Such signals can only be expected for SNe in the nearby galaxy (e.g., \citealt{derkacy2024,burrow2024}). Besides, the local environments of SNe Ia at low redshifts would have been polluted by multiple cycles of Pop I/II SNe. 
Above all, it is still an open question to distinguish Pop III SNe Ia from their Pop I/II counterparts based on their observational characteristics. This is a subject that deserves further investigation.}

\section{Summary and Conclusion}
\label{sec:5}

In this {work}, we investigated the formation of SNe Ia from Pop III binaries and discussed its implications for the observations. The SD channel for SN Ia is studied assuming the CEW model. We simulated the WD mass accumulation processes through detailed binary evolution calculations and derived the initial parameter spaces that can produce SNe Ia successfully. As for the DD channel, the approximately power-low DTD of \citet{wangb2012} is adopted. Then, we calculate the SN Ia rate as a function of redshift with different input parameters, based on theoretical models of CSFH ({S19} and LB20) and IMF (bottom-heavy {and} top-heavy IMFs). 


The SN Ia rate from Pop III stars peaks at $z\sim 6.5$ for {S19} model and at $z\sim8$ for LB20 model. Due to its large star formation rate at $z\sim 7-12$, {S19} model predicts more SNe Ia than the LB20 model with other parameters fixed. We found that the rates depend strongly on the IMFs, in particular, on the upper limit of stellar mass of a top-heavy IMF. In the optimistic model, i.e., top-heavy IMF of $M_{\rm up}=10$ M$_\odot$ ({about} $90\%$ star masses are in the range of $1-10$ M$_\odot$), the SN Ia rate of Pop III stars is {comparable to or even higher than} that of the Pop I/II stars at redshift $z\sim 5-6$. 
We also calculate the expected number for the JWST with a survey area of $300\;\rm arcmin^2$ and survey time of $3\;\rm yr$. {Ideally, we may expect $\sim 1$ Pop III SN Ia detected by JWST up to $z\sim 4$ and $\sim 2$ Pop III SNe Ia up to $z\sim 5$. We found $\sim 8$ Pop III SNe Ia may exist in the FLARE field at redshifts of $5-10$.} 
{SNe Ia at redshifts up to $z\sim 10$ can be detected if the JWST survey can reach a limit magnitude down to $\sim 28.5$ mag \citep{wangl2017}.}
For $z\lesssim2.5$, the number of SNe Ia produced from Pop I/II stars is higher than those from Pop III stars by about two orders of magnitude ($\sim 400$) in the {S19} model. It suggests that for a complete redshift limited survey more than $\sim 400$ SNe Ia found at low redshift may contain one Pop III SN Ia event. {Unfortunately, we still do not know the obvious characteristic of the Pop III SN Ia compared with Pop I/II SN Ia. This issue is worthy of further research. }



Due to the unknown observational constraints on the Pop III IMF, the mass spectrum of Pop III stars are highly uncertain. We note that our results strongly depends on the IMF, more specifically, the fraction of star masses in the range of $1-10$ M$_\odot$. In fact, many recent numerical simulations suggest the IMF shifts towards low-mass stars (e.g., \citealt{jaura2022,prole2022}). However, simulations of primordial stars are still quite limited, on account of the spatial resolution, the time span, and the details of the numerical implementation \citep{klessen2023}. Therefore, our results proposed an upper limit of Pop III SN Ia rate. On the other side, the detection of SN Ia at high redshift in the future may confirm that a significant part of masses are in low-mass stars for Pop III stars. 

\section*{Acknowledgements}
{We cordially thank the anonymous referee, whose comments helped to improve the quality of the manuscript a lot.} We thank the helpful discussion about the CEW model with Xiangcun Meng. We also thank Filippo Santoliquido, Eniko Regos, and Jozsef Vinko for sharing the code in calculating SN Ia rate. ZWL would like to thank the continuous encouragement and kindly suggestions from Dihong Gong. This work is supported by the National Key R$\&$D Program of China (grant Nos. 2021YFA1600403, 2021YFA1600400), the Natural Science Foundation of China (grant Nos. 12125303, 12288102, 12090040/3, 12103086, 12273105, 11703081, 11422324, 12073070), the Natural Science Foundation of Yunnan Province (No. 202201BC070003) and the Yunnan Revitalization Talent Support Program–Science $\&$ Technology Champion Project (No. 202305AB350003), the Key Research Program of Frontier Sciences of CAS (No. ZDBS-LY-7005), Yunnan Fundamental Research Projects (grant Nos. 202301AT070314, 202101AU070276, 202101AV070001), and the International Centre of Supernovae, Yunnan Key Laboratory (No. 202302AN360001). LW is grateful to NSF for the grant AST 1813825 which partially support his research. We also acknowledge the science research grant from the China Manned Space Project with Nos. CMS-CSST-2021-A10 and CMS-CSST-2021-A08. The authors gratefully acknowledge the “PHOENIX Supercomputing Platform” jointly operated by the Binary Population Synthesis Group and the Stellar Astrophysics Group at Yunnan Observatories, Chinese Academy of Sciences. 

\software{\texttt{MESA} (v12115; \citealt{paxton2011,paxton2013,paxton2015,paxton2018,paxton2019})}

\appendix 
\section{Typical timescale of WD progenitors}
\label{app:A}
{In the first step, we try to find the typical progenitor mass for a given WD. 
It is neccessary to recall the formation channel of detached WD+MS binaries, i.e., the progenitor binaries of SNe Ia in single degenerate channel. The formation of WD+MS binaries involves two basic binary interaction processes, namely stable Roche lobe overflow (RLOF) and CE ejection. If the WD is born from RLOF, the orbital periods of WD+MS binaries are generally larger than $\sim 100\;\rm d$ \citep{willems2004,lizw2023}. Then we could expect only a small part of WD+MS binaries from RLOF channel make a contribution to SNe Ia. If the WD+MS binaries are produced from CE ejection processes. The orbital separation would shrink by $\sim 10-50$ times, and the typical orbital periods are in the range of $0.1-100\;\rm d$ \citep{willems2004}, which well covers the parameter space that leads to the SNe Ia. Therefore, we assume that all WD+MS binaries that produces SNe Ia have experienced the CE ejection processes.  }

\begin{figure}
    \centering
    \includegraphics[width=0.5\columnwidth]{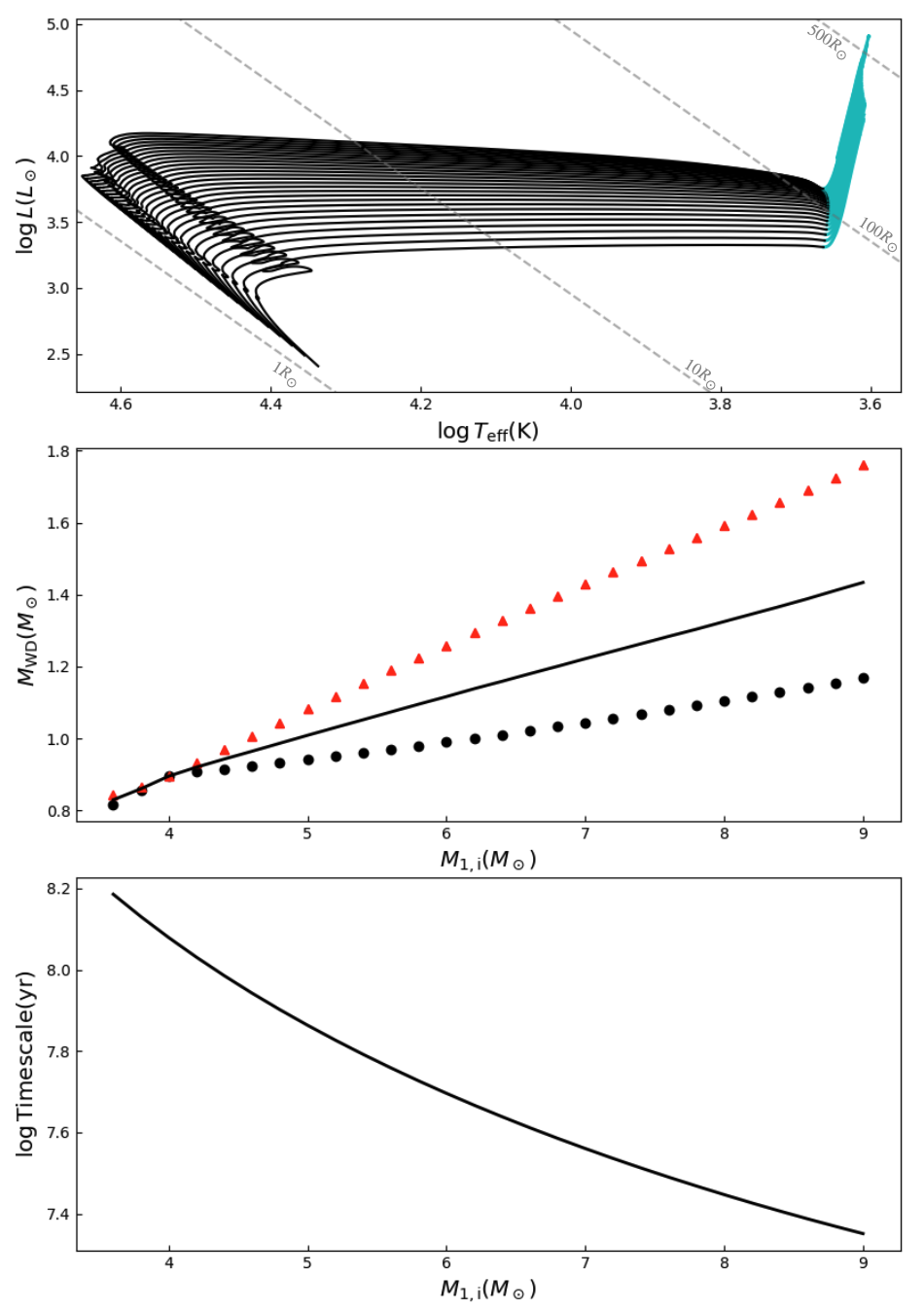}
    \caption{\textbf{Upper panel:} The evolutionary tracks of stars with masses from $3.6$ to $9.0M_\odot$ in a step of $0.2M_\odot$. The cyan lines represents the stages where the thick convective envelope have been developed, analogous to the AGB stage of Pop I/II stars. \textbf{Middle panel:} The helium core mass for stars with thick convective envelopes (cyan lines at the upper panel). The dredge-up processes may happen in the convective regions and reducing the helium core mass. The minimum and maximum helium core masses at the AGB stages are shown in black circles and red rectangles, respectively. The black solid line represents the logarithm interpolation between the maximum and minimum values of helium core masses, which gives the typical WD progenitor mass for a given WD. \textbf{Lower panel: }The typical evolutionary timescale of the progenitor stars. The timescales are calculated from ZAMS stages to the bottom of AGB stage (the lower ends of the cyan lines). }
    \label{fig:A1}
\end{figure}

{We evolve several evolutionary tracks of stars with masses from $3.6$ to $9.0M_\odot$ in a step of $0.2M_\odot$, as shown in the upper panel of Figure~\ref{fig:A1}. Here we neglect the stellar winds during the evolution (The effect of stellar wind on the final WD mass can be referred to \citealt{lawlor2023}). We assume that the CE phase happens as the primary star develops a thick convective envelope (at giant branch). The stages of stars with considerable convective envelopes are shown in cyan lines. The deep convective region may penetrate into the helium core, resulting in the so-called dredge-up process. During the dredge-up phase, the helium core mass decreases with the stellar evolution. The minimum and maximum helium core masses at this stage are shown in black circles and red rectangles of the middle panel, respectively. The resulted WD masses after the CE ejection processes are within the region between the red rectangles and black circles. We adopt the logarithm interpolation method to approximately give the typical progenitor mass for a given WD, as shown in the black solid line of the middle panel. The typical evolutionary timescale of the progenitors as a function of initial progenitor masses is shown in the lower panel, where the timescale includes the MS lifetime and the He-shell burning lifetime (from ZAMS stages to the lower ends of the cyan lines). }

\newpage

\section{SN Ia rates with different CSFHs}
\label{app:B}
\begin{figure*}
	\begin{minipage}[t]{0.5\textwidth}
		\centering
		\includegraphics[width=\textwidth]{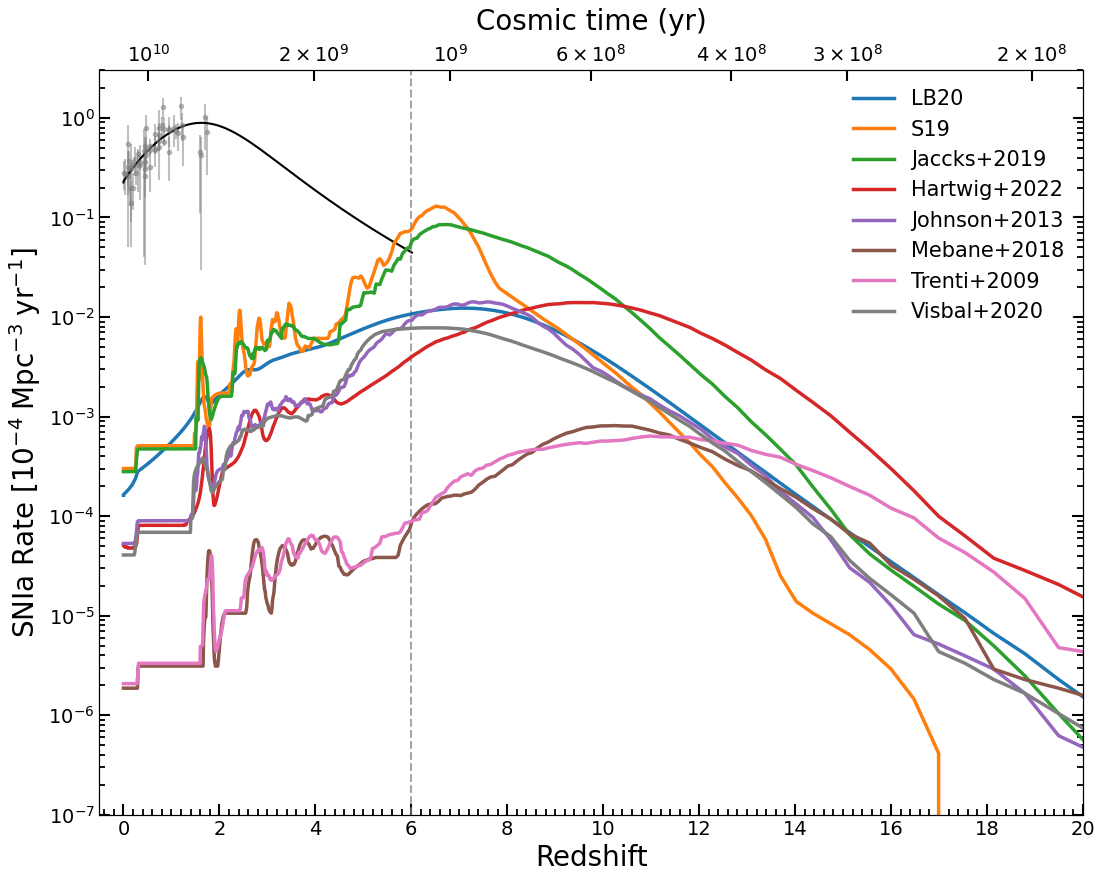}
	\end{minipage}
	\begin{minipage}[t]{0.5\textwidth}
		\centering
		\includegraphics[width=\textwidth]{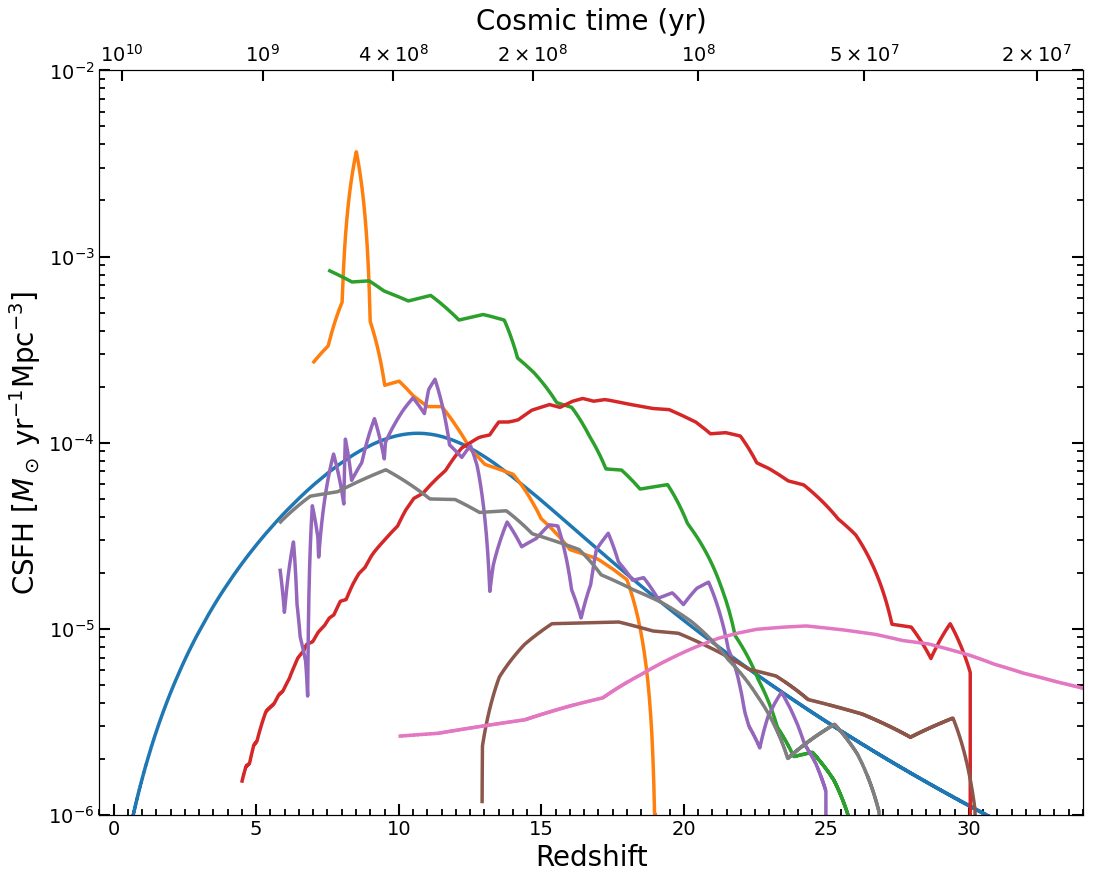}
	\end{minipage}
	\caption{Left panel: The influence of CSFHs on SN Ia rates. The Pop III SN Ia rates are obtained based on the optimistic model, i.e., $f_{\rm b}=1$ and top-heavy function of $M_{\rm up}=10M_\odot$. Right panel: The CSFHs for Pop III stars in several studies from \citet{liub2020,sarmento2019,jaacks2019,hartwig2022,johnson2013,mebane2018,trenti2009,visbal2020}.}
    \label{fig:B1}
\end{figure*}

{We calculate the influence of CSFHs taken from several previous works on the SN Ia rates, as shown in Figure~\ref{fig:B1}. The Pop III SN Ia rates are obtained based on the optimistic model, i.e., $f_{\rm b}=1$ and top-heavy function of $M_{\rm up}=10M_\odot$. The rapid increase of SFRH at $z\sim 8.5$ of \citet{sarmento2019} is caused by a significant number of halos crossing the density threshold for star formation before the reionization (see also \citealt{sarmento2018}). We see that the results of SN Ia rates would vary by about $\sim 2$ orders of magnitude for different CSFHs. }

\newpage

\bibliographystyle{aasjournal}
\bibliography{refs}

\begin{thebibliography}{}
\expandafter\ifx\csname natexlab\endcsname\relax\def\natexlab#1{#1}\fi
\providecommand{\url}[1]{\href{#1}{#1}}
\providecommand{\dodoi}[1]{doi:~\href{http://doi.org/#1}{\nolinkurl{#1}}}
\providecommand{\doeprint}[1]{\href{http://ascl.net/#1}{\nolinkurl{http://ascl.net/#1}}}
\providecommand{\doarXiv}[1]{\href{https://arxiv.org/abs/#1}{\nolinkurl{https://arxiv.org/abs/#1}}}

\bibitem[{{Abel} {et~al.}(2002){Abel}, {Bryan}, \& {Norman}}]{abel2002}
{Abel}, T., {Bryan}, G.~L., \& {Norman}, M.~L. 2002, Science, 295, 93,
  \dodoi{10.1126/science.295.5552.93}

\bibitem[{{Abt}(1983)}]{abt1983}
{Abt}, H.~A. 1983, \araa, 21, 343, \dodoi{10.1146/annurev.aa.21.090183.002015}

\bibitem[{{Aryan} {et~al.}(2023){Aryan}, {Pandey}, {Gupta}, \&
  {Ror}}]{aryan2023}
{Aryan}, A., {Pandey}, S.~B., {Gupta}, R., \& {Ror}, A.~K. 2023, \mnras, 521,
  L17, \dodoi{10.1093/mnrasl/slad020}

\bibitem[{{Bazin} {et~al.}(2009){Bazin}, {Palanque-Delabrouille}, {Rich},
  {Ruhlmann-Kleider}, {Aubourg}, {Le Guillou}, {Astier}, {Balland}, {Basa},
  {Carlberg}, {Conley}, {Fouchez}, {Guy}, {Hardin}, {Hook}, {Howell}, {Pain},
  {Perrett}, {Pritchet}, {Regnault}, {Sullivan}, {Antilogus}, {Arsenijevic},
  {Baumont}, {Fabbro}, {Le Du}, {Lidman}, {Mouchet}, {Mour{\~a}o}, \&
  {Walker}}]{bazin2009}
{Bazin}, G., {Palanque-Delabrouille}, N., {Rich}, J., {et~al.} 2009, \aap, 499,
  653, \dodoi{10.1051/0004-6361/200911847}

\bibitem[{{Bromm} {et~al.}(2002){Bromm}, {Coppi}, \& {Larson}}]{bromm2002}
{Bromm}, V., {Coppi}, P.~S., \& {Larson}, R.~B. 2002, \apj, 564, 23,
  \dodoi{10.1086/323947}

\bibitem[{{Bullock} \& {Boylan-Kolchin}(2017)}]{bullock2017}
{Bullock}, J.~S., \& {Boylan-Kolchin}, M. 2017, \araa, 55, 343,
  \dodoi{10.1146/annurev-astro-091916-055313}

\bibitem[{{Burrow} {et~al.}(2024){Burrow}, {Baron}, {Burns}, {Hsiao}, {Lu},
  {Ashall}, {Brown}, {DerKacy}, {Folatelli}, {Galbany}, {Hoeflich},
  {Krisciunas}, {Morrell}, {Phillips}, {Shappee}, {Stritzinger}, \&
  {Suntzeff}}]{burrow2024}
{Burrow}, A., {Baron}, E., {Burns}, C.~R., {et~al.} 2024, arXiv e-prints,
  arXiv:2404.04724, \dodoi{10.48550/arXiv.2404.04724}

\bibitem[{{Cassisi} \& {Castellani}(1993)}]{cassisi1993}
{Cassisi}, S., \& {Castellani}, V. 1993, \apjs, 88, 509, \dodoi{10.1086/191831}

\bibitem[{{Castellano} {et~al.}(2023){Castellano}, {Fontana}, {Treu}, {Merlin},
  {Santini}, {Bergamini}, {Grillo}, {Rosati}, {Acebron}, {Leethochawalit},
  {Paris}, {Bonchi}, {Belfiori}, {Calabr{\`o}}, {Correnti}, {Nonino},
  {Polenta}, {Trenti}, {Boyett}, {Brammer}, {Broadhurst}, {Caminha}, {Chen},
  {Filippenko}, {Fortuni}, {Glazebrook}, {Mascia}, {Mason}, {Menci},
  {Meneghetti}, {Mercurio}, {Metha}, {Morishita}, {Nanayakkara}, {Pentericci},
  {Roberts-Borsani}, {Roy}, {Vanzella}, {Vulcani}, {Yang}, \&
  {Wang}}]{castellano2023}
{Castellano}, M., {Fontana}, A., {Treu}, T., {et~al.} 2023, \apjl, 948, L14,
  \dodoi{10.3847/2041-8213/accea5}

\bibitem[{Chen {et~al.}(2019)Chen, Woods, Yungelson, Piersanti, Gilfanov, \&
  Han}]{chenh2019}
Chen, H.-L., Woods, T.~E., Yungelson, L.~R., {et~al.} 2019, Monthly Notices of
  the Royal Astronomical Society, 490, 1678–1692,
  \dodoi{10.1093/mnras/stz2644}

\bibitem[{{Chen} {et~al.}(2014){Chen}, {Herwig}, {Denissenkov}, \&
  {Paxton}}]{chenm2014}
{Chen}, M.~C., {Herwig}, F., {Denissenkov}, P.~A., \& {Paxton}, B. 2014,
  \mnras, 440, 1274, \dodoi{10.1093/mnras/stu108}

\bibitem[{{Chen} {et~al.}(2011){Chen}, {Han}, \& {Tout}}]{chenx2011}
{Chen}, X., {Han}, Z., \& {Tout}, C.~A. 2011, \apjl, 735, L31,
  \dodoi{10.1088/2041-8205/735/2/L31}

\bibitem[{{Claeys} {et~al.}(2014){Claeys}, {Pols}, {Izzard}, {Vink}, \&
  {Verbunt}}]{claeys2014}
{Claeys}, J.~S.~W., {Pols}, O.~R., {Izzard}, R.~G., {Vink}, J., \& {Verbunt},
  F.~W.~M. 2014, \aap, 563, A83, \dodoi{10.1051/0004-6361/201322714}

\bibitem[{{DerKacy} {et~al.}(2024){DerKacy}, {Ashall}, {Hoeflich}, {Baron},
  {Shahbandeh}, {Shappee}, {Andrews}, {Baade}, {Balangan}, {Bostroem}, {Brown},
  {Burns}, {Burrow}, {Cikota}, {de Jaeger}, {Do}, {Dong}, {Dominguez}, {Fox},
  {Galbany}, {Hoang}, {Hsiao}, {Janzen}, {Jencson}, {Krisciunas}, {Kumar},
  {Lu}, {Lundquist}, {Mera Evans}, {Maund}, {Mazzali}, {Medler}, {Meza
  Retamal}, {Morrell}, {Patat}, {Pearson}, {Phillips}, {Shrestha}, {Stangl},
  {Stevens}, {Stritzinger}, {Suntzeff}, {Telesco}, {Tucker}, {Valenti}, {Wang},
  \& {Yang}}]{derkacy2024}
{DerKacy}, J.~M., {Ashall}, C., {Hoeflich}, P., {et~al.} 2024, \apj, 961, 187,
  \dodoi{10.3847/1538-4357/ad0b7b}

\bibitem[{{DES Collaboration} {et~al.}(2024){DES Collaboration}, {Abbott},
  {Acevedo}, {Aguena}, {Alarcon}, {Allam}, {Alves}, {Amon}, {Andrade-Oliveira},
  {Annis}, {Armstrong}, {Asorey}, {Avila}, {Bacon}, {Bassett}, {Bechtol},
  {Bernardinelli}, {Bernstein}, {Bertin}, {Blazek}, {Bocquet}, {Brooks},
  {Brout}, {Buckley-Geer}, {Burke}, {Camacho}, {Camilleri}, {Campos}, {Carnero
  Rosell}, {Carollo}, {Carr}, {Carretero}, {Castander}, {Cawthon}, {Chang},
  {Chen}, {Choi}, {Conselice}, {Costanzi}, {da Costa}, {Crocce}, {Davis},
  {DePoy}, {Desai}, {Diehl}, {Dixon}, {Dodelson}, {Doel}, {Doux},
  {Drlica-Wagner}, {Elvin-Poole}, {Everett}, {Ferrero}, {Fert{\'e}},
  {Flaugher}, {Foley}, {Fosalba}, {Friedel}, {Frieman}, {Frohmaier}, {Galbany},
  {Garc{\'\i}a-Bellido}, {Gatti}, {Gaztanaga}, {Giannini}, {Glazebrook},
  {Graur}, {Gruen}, {Gruendl}, {Gutierrez}, {Hartley}, {Herner}, {Hinton},
  {Hollowood}, {Honscheid}, {Huterer}, {Jain}, {James}, {Jeffrey}, {Kelsey},
  {Kent}, {Kessler}, {Kim}, {Kirshner}, {Kovacs}, {Kuehn}, {Lahav}, {Lee},
  {Lee}, {Lewis}, {Li}, {Lidman}, {Lin}, {Marshall}, {Martini},
  {Mena-Fern{\'a}ndez}, {Menanteau}, {Miquel}, {Mohr}, {Mould}, {Muir},
  {M{\"o}ller}, {Neilsen}, {Nichol}, {Nugent}, {Ogando}, {Palmese}, {Pan},
  {Paterno}, {Percival}, {Pereira}, {Pieres}, {Plazas Malag{\'o}n}, {Popovic},
  {Porredon}, {Prat}, {Qu}, {Raveri}, {Rodr{\'\i}guez-Monroy}, {Romer},
  {Roodman}, {Rose}, {Sako}, {Sanchez}, {Sanchez Cid}, {Schubnell}, {Scolnic},
  {Sevilla-Noarbe}, {Shah}, {Allyn. Smith}, {Smith}, {Soares-Santos},
  {Suchyta}, {Sullivan}, {Suntzeff}, {Swanson}, {S{\'a}nchez}, {Tarle},
  {Taylor}, {Thomas}, {To}, {Toy}, {Troxel}, {Tucker}, {Tucker}, {Uddin},
  {Vincenzi}, {Walker}, {Weaverdyck}, {Wechsler}, {Weller}, {Wester},
  {Wiseman}, {Yamamoto}, {Yuan}, {Zhang}, \& {Zhang}}]{des2024}
{DES Collaboration}, {Abbott}, T.~M.~C., {Acevedo}, M., {et~al.} 2024, arXiv
  e-prints, arXiv:2401.02929, \dodoi{10.48550/arXiv.2401.02929}

\bibitem[{{Gardner} {et~al.}(2006){Gardner}, {Mather}, {Clampin}, {Doyon},
  {Greenhouse}, {Hammel}, {Hutchings}, {Jakobsen}, {Lilly}, {Long}, {Lunine},
  {McCaughrean}, {Mountain}, {Nella}, {Rieke}, {Rieke}, {Rix}, {Smith},
  {Sonneborn}, {Stiavelli}, {Stockman}, {Windhorst}, \& {Wright}}]{gardner2006}
{Gardner}, J.~P., {Mather}, J.~C., {Clampin}, M., {et~al.} 2006, \ssr, 123,
  485, \dodoi{10.1007/s11214-006-8315-7}

\bibitem[{{Ge} {et~al.}(2023){Ge}, {Tout}, {Chen}, {Sarkar}, {Walton}, \&
  {Han}}]{geh2023}
{Ge}, H., {Tout}, C.~A., {Chen}, X., {et~al.} 2023, \apj, 945, 7,
  \dodoi{10.3847/1538-4357/acb7e9}

\bibitem[{{Greggio} \& {Renzini}(1983)}]{greggio1983}
{Greggio}, L., \& {Renzini}, A. 1983, \aap, 118, 217

\bibitem[{{Greif} {et~al.}(2012){Greif}, {Bromm}, {Clark}, {Glover}, {Smith},
  {Klessen}, {Yoshida}, \& {Springel}}]{greif2012}
{Greif}, T.~H., {Bromm}, V., {Clark}, P.~C., {et~al.} 2012, \mnras, 424, 399,
  \dodoi{10.1111/j.1365-2966.2012.21212.x}

\bibitem[{{Greif} {et~al.}(2011){Greif}, {Springel}, {White}, {Glover},
  {Clark}, {Smith}, {Klessen}, \& {Bromm}}]{greif2011}
{Greif}, T.~H., {Springel}, V., {White}, S. D.~M., {et~al.} 2011, \apj, 737,
  75, \dodoi{10.1088/0004-637X/737/2/75}

\bibitem[{{Guo} {et~al.}(2022){Guo}, {Liu}, {Wang}, {Wang}, {Zhang}, {Ji},
  {Han}, \& {Chen}}]{guoy2022}
{Guo}, Y., {Liu}, C., {Wang}, L., {et~al.} 2022, \aap, 667, A44,
  \dodoi{10.1051/0004-6361/202244300}

\bibitem[{{Hachisu} {et~al.}(1996){Hachisu}, {Kato}, \& {Nomoto}}]{hachisu1996}
{Hachisu}, I., {Kato}, M., \& {Nomoto}, K. 1996, \apjl, 470, L97,
  \dodoi{10.1086/310303}

\bibitem[{{Hachisu} {et~al.}(1999){Hachisu}, {Kato}, {Nomoto}, \&
  {Umeda}}]{hachisu1999}
{Hachisu}, I., {Kato}, M., {Nomoto}, K., \& {Umeda}, H. 1999, \apj, 519, 314,
  \dodoi{10.1086/307370}

\bibitem[{{Han} \& {Podsiadlowski}(2004)}]{han2004}
{Han}, Z., \& {Podsiadlowski}, P. 2004, \mnras, 350, 1301,
  \dodoi{10.1111/j.1365-2966.2004.07713.x}

\bibitem[{{Hartwig} {et~al.}(2022){Hartwig}, {Magg}, {Chen}, {Tarumi}, {Bromm},
  {Glover}, {Ji}, {Klessen}, {Latif}, {Volonteri}, \& {Yoshida}}]{hartwig2022}
{Hartwig}, T., {Magg}, M., {Chen}, L.-H., {et~al.} 2022, \apj, 936, 45,
  \dodoi{10.3847/1538-4357/ac7150}

\bibitem[{{Heger} \& {Woosley}(2002)}]{heger2002}
{Heger}, A., \& {Woosley}, S.~E. 2002, \apj, 567, 532, \dodoi{10.1086/338487}

\bibitem[{{Herwig}(2000)}]{herwig2000}
{Herwig}, F. 2000, \aap, 360, 952, \dodoi{10.48550/arXiv.astro-ph/0007139}

\bibitem[{{Hirano} {et~al.}(2015){Hirano}, {Hosokawa}, {Yoshida}, {Omukai}, \&
  {Yorke}}]{hirano2015}
{Hirano}, S., {Hosokawa}, T., {Yoshida}, N., {Omukai}, K., \& {Yorke}, H.~W.
  2015, \mnras, 448, 568, \dodoi{10.1093/mnras/stv044}

\bibitem[{{Hirano} {et~al.}(2014){Hirano}, {Hosokawa}, {Yoshida}, {Umeda},
  {Omukai}, {Chiaki}, \& {Yorke}}]{hirano2014}
{Hirano}, S., {Hosokawa}, T., {Yoshida}, N., {et~al.} 2014, \apj, 781, 60,
  \dodoi{10.1088/0004-637X/781/2/60}

\bibitem[{{Hopkins} \& {Beacom}(2006)}]{hopkins2006}
{Hopkins}, A.~M., \& {Beacom}, J.~F. 2006, \apj, 651, 142,
  \dodoi{10.1086/506610}

\bibitem[{{Hosokawa} {et~al.}(2016){Hosokawa}, {Hirano}, {Kuiper}, {Yorke},
  {Omukai}, \& {Yoshida}}]{hosokawa2016}
{Hosokawa}, T., {Hirano}, S., {Kuiper}, R., {et~al.} 2016, \apj, 824, 119,
  \dodoi{10.3847/0004-637X/824/2/119}

\bibitem[{{Hoyle} \& {Fowler}(1960)}]{hoyle1960}
{Hoyle}, F., \& {Fowler}, W.~A. 1960, \apj, 132, 565, \dodoi{10.1086/146963}

\bibitem[{{Hurley} {et~al.}(2002){Hurley}, {Tout}, \& {Pols}}]{Hurley02}
{Hurley}, J.~R., {Tout}, C.~A., \& {Pols}, O.~R. 2002, \mnras, 329, 897,
  \dodoi{10.1046/j.1365-8711.2002.05038.x}

\bibitem[{{Iben} \& {Tutukov}(1984)}]{iben1984}
{Iben}, I., J., \& {Tutukov}, A.~V. 1984, \apjs, 54, 335,
  \dodoi{10.1086/190932}

\bibitem[{{Jaacks} {et~al.}(2019){Jaacks}, {Finkelstein}, \&
  {Bromm}}]{jaacks2019}
{Jaacks}, J., {Finkelstein}, S.~L., \& {Bromm}, V. 2019, \mnras, 488, 2202,
  \dodoi{10.1093/mnras/stz1529}

\bibitem[{{Jaura} {et~al.}(2022){Jaura}, {Glover}, {Wollenberg}, {Klessen},
  {Geen}, \& {Haemmerl{\'e}}}]{jaura2022}
{Jaura}, O., {Glover}, S. C.~O., {Wollenberg}, K. M.~J., {et~al.} 2022, \mnras,
  512, 116, \dodoi{10.1093/mnras/stac487}

\bibitem[{{Johnson} {et~al.}(2013){Johnson}, {Dalla Vecchia}, \&
  {Khochfar}}]{johnson2013}
{Johnson}, J.~L., {Dalla Vecchia}, C., \& {Khochfar}, S. 2013, \mnras, 428,
  1857, \dodoi{10.1093/mnras/sts011}

\bibitem[{{Kato} \& {Hachisu}(2004)}]{kato2004}
{Kato}, M., \& {Hachisu}, I. 2004, \apjl, 613, L129, \dodoi{10.1086/425249}

\bibitem[{{Klessen} \& {Glover}(2023)}]{klessen2023}
{Klessen}, R.~S., \& {Glover}, S. C.~O. 2023, \araa, 61, 65,
  \dodoi{10.1146/annurev-astro-071221-053453}

\bibitem[{{Kolb} \& {Ritter}(1990)}]{kolb1990}
{Kolb}, U., \& {Ritter}, H. 1990, \aap, 236, 385

\bibitem[{{Kroupa} {et~al.}(1993){Kroupa}, {Tout}, \& {Gilmore}}]{kroupa1993}
{Kroupa}, P., {Tout}, C.~A., \& {Gilmore}, G. 1993, \mnras, 262, 545,
  \dodoi{10.1093/mnras/262.3.545}

\bibitem[{{Latif} {et~al.}(2022){Latif}, {Whalen}, \& {Khochfar}}]{latif2022}
{Latif}, M.~A., {Whalen}, D., \& {Khochfar}, S. 2022, \apj, 925, 28,
  \dodoi{10.3847/1538-4357/ac3916}

\bibitem[{{Lawlor} \& {MacDonald}(2023)}]{lawlor2023}
{Lawlor}, T.~M., \& {MacDonald}, J. 2023, \mnras, 525, 4700,
  \dodoi{10.1093/mnras/stad2582}

\bibitem[{{Lawlor} {et~al.}(2008){Lawlor}, {Young}, {Johnson}, \&
  {MacDonald}}]{lawlor2008}
{Lawlor}, T.~M., {Young}, T.~R., {Johnson}, T.~A., \& {MacDonald}, J. 2008,
  \mnras, 384, 1533, \dodoi{10.1111/j.1365-2966.2007.12810.x}

\bibitem[{{Leonard}(2007)}]{leonard2007}
{Leonard}, D.~C. 2007, \apj, 670, 1275, \dodoi{10.1086/522367}

\bibitem[{{Leung} \& {Nomoto}(2018)}]{leung2018}
{Leung}, S.-C., \& {Nomoto}, K. 2018, \apj, 861, 143,
  \dodoi{10.3847/1538-4357/aac2df}

\bibitem[{{Leung} \& {Nomoto}(2020)}]{leung2020}
---. 2020, \apj, 900, 54, \dodoi{10.3847/1538-4357/aba1e3}

\bibitem[{{Li} {et~al.}(2019){Li}, {Chen}, {Chen}, \& {Han}}]{lizw2019}
{Li}, Z., {Chen}, X., {Chen}, H.-L., \& {Han}, Z. 2019, \apj, 871, 148,
  \dodoi{10.3847/1538-4357/aaf9a1}

\bibitem[{{Li} {et~al.}(2023){Li}, {Chen}, {Ge}, {Chen}, \& {Han}}]{lizw2023}
{Li}, Z., {Chen}, X., {Ge}, H., {Chen}, H.-L., \& {Han}, Z. 2023, \aap, 669,
  A82, \dodoi{10.1051/0004-6361/202243893}

\bibitem[{{Liu} \& {Bromm}(2020)}]{liub2020}
{Liu}, B., \& {Bromm}, V. 2020, \mnras, 497, 2839,
  \dodoi{10.1093/mnras/staa2143}

\bibitem[{{Liu} {et~al.}(2018){Liu}, {Wang}, \& {Han}}]{liud2018}
{Liu}, D., {Wang}, B., \& {Han}, Z. 2018, \mnras, 473, 5352,
  \dodoi{10.1093/mnras/stx2756}

\bibitem[{{Liu} {et~al.}(2023){Liu}, {R{\"o}pke}, \& {Han}}]{liuz2023}
{Liu}, Z.-W., {R{\"o}pke}, F.~K., \& {Han}, Z. 2023, Research in Astronomy and
  Astrophysics, 23, 082001, \dodoi{10.1088/1674-4527/acd89e}

\bibitem[{{LSST Science Collaboration} {et~al.}(2009){LSST Science
  Collaboration}, {Abell}, {Allison}, {Anderson}, {Andrew}, {Angel}, {Armus},
  {Arnett}, {Asztalos}, {Axelrod}, \& et~al.}]{lsst2009}
{LSST Science Collaboration}, {Abell}, P.~A., {Allison}, J., {et~al.} 2009,
  arXiv e-prints.
\newblock \doarXiv{0912.0201}

\bibitem[{{Lu} {et~al.}(2023){Lu}, {Fuller}, {Quataert}, \&
  {Bonnerot}}]{luw2023}
{Lu}, W., {Fuller}, J., {Quataert}, E., \& {Bonnerot}, C. 2023, \mnras, 519,
  1409, \dodoi{10.1093/mnras/stac3621}

\bibitem[{{Madau} \& {Dickinson}(2014)}]{madau2014}
{Madau}, P., \& {Dickinson}, M. 2014, \araa, 52, 415,
  \dodoi{10.1146/annurev-astro-081811-125615}

\bibitem[{{Madau} \& {Fragos}(2017)}]{madau2017}
{Madau}, P., \& {Fragos}, T. 2017, \apj, 840, 39,
  \dodoi{10.3847/1538-4357/aa6af9}

\bibitem[{{Maguire} {et~al.}(2016){Maguire}, {Taubenberger}, {Sullivan}, \&
  {Mazzali}}]{maguire2016}
{Maguire}, K., {Taubenberger}, S., {Sullivan}, M., \& {Mazzali}, P.~A. 2016,
  \mnras, 457, 3254, \dodoi{10.1093/mnras/stv2991}

\bibitem[{{Maiolino} {et~al.}(2023){Maiolino}, {Uebler}, {Perna}, {Scholtz},
  {D'Eugenio}, {Witten}, {Laporte}, {Witstok}, {Carniani}, {Tacchella},
  {Baker}, {Arribas}, {Nakajima}, {Eisenstein}, {Bunker}, {Charlot}, {Cresci},
  {Curti}, {Curtis-Lake}, {de Graaff}, {Ji}, {Johnson}, {Kumari}, {Looser},
  {Maseda}, {Robertson}, {Rodriguez Del Pino}, {Sandles}, {Simmonds}, {Smit},
  {Sun}, {Venturi}, {Williams}, \& {Willmer}}]{maiolino2023}
{Maiolino}, R., {Uebler}, H., {Perna}, M., {et~al.} 2023, arXiv e-prints,
  arXiv:2306.00953, \dodoi{10.48550/arXiv.2306.00953}

\bibitem[{{Maoz} \& {Graur}(2017)}]{maoz2017}
{Maoz}, D., \& {Graur}, O. 2017, \apj, 848, 25,
  \dodoi{10.3847/1538-4357/aa8b6e}

\bibitem[{{Marigo} {et~al.}(2001){Marigo}, {Girardi}, {Chiosi}, \&
  {Wood}}]{marigo2001}
{Marigo}, P., {Girardi}, L., {Chiosi}, C., \& {Wood}, P.~R. 2001, \aap, 371,
  152, \dodoi{10.1051/0004-6361:20010309}

\bibitem[{{Matteucci} \& {Greggio}(1986)}]{matteucci1986}
{Matteucci}, F., \& {Greggio}, L. 1986, \aap, 154, 279

\bibitem[{{Mazeh} {et~al.}(1992){Mazeh}, {Goldberg}, {Duquennoy}, \&
  {Mayor}}]{mazeh1992}
{Mazeh}, T., {Goldberg}, D., {Duquennoy}, A., \& {Mayor}, M. 1992, \apj, 401,
  265, \dodoi{10.1086/172058}

\bibitem[{{Mebane} {et~al.}(2018){Mebane}, {Mirocha}, \&
  {Furlanetto}}]{mebane2018}
{Mebane}, R.~H., {Mirocha}, J., \& {Furlanetto}, S.~R. 2018, \mnras, 479, 4544,
  \dodoi{10.1093/mnras/sty1833}

\bibitem[{{Meng} {et~al.}(2008){Meng}, {Chen}, \& {Han}}]{mengx2008}
{Meng}, X., {Chen}, X., \& {Han}, Z. 2008, \aap, 487, 625,
  \dodoi{10.1051/0004-6361:20078841}

\bibitem[{{Meng} {et~al.}(2009){Meng}, {Chen}, \& {Han}}]{meng2009}
---. 2009, \mnras, 395, 2103, \dodoi{10.1111/j.1365-2966.2009.14636.x}

\bibitem[{{Meng} \& {Podsiadlowski}(2014)}]{mengx2014}
{Meng}, X., \& {Podsiadlowski}, P. 2014, \apjl, 789, L45,
  \dodoi{10.1088/2041-8205/789/2/L45}

\bibitem[{{Meng} \& {Podsiadlowski}(2017)}]{meng2017}
---. 2017, \mnras, 469, 4763, \dodoi{10.1093/mnras/stx1137}

\bibitem[{{Moe} \& {Di Stefano}(2017)}]{moe2017}
{Moe}, M., \& {Di Stefano}, R. 2017, \apjs, 230, 15,
  \dodoi{10.3847/1538-4365/aa6fb6}

\bibitem[{{Moriya} {et~al.}(2019){Moriya}, {Wong}, {Koyama}, {Tanaka}, {Oguri},
  {Hilbert}, \& {Nomoto}}]{moriya2019}
{Moriya}, T.~J., {Wong}, K.~C., {Koyama}, Y., {et~al.} 2019, \pasj, 71, 59,
  \dodoi{10.1093/pasj/psz035}

\bibitem[{{Nomoto} {et~al.}(1984){Nomoto}, {Thielemann}, \&
  {Yokoi}}]{nomoto1984}
{Nomoto}, K., {Thielemann}, F.~K., \& {Yokoi}, K. 1984, \apj, 286, 644,
  \dodoi{10.1086/162639}

\bibitem[{{Omukai} \& {Nishi}(1998)}]{omukai1998}
{Omukai}, K., \& {Nishi}, R. 1998, \apj, 508, 141, \dodoi{10.1086/306395}

\bibitem[{{Paxton} {et~al.}(2011){Paxton}, {Bildsten}, {Dotter}, {Herwig},
  {Lesaffre}, \& {Timmes}}]{paxton2011}
{Paxton}, B., {Bildsten}, L., {Dotter}, A., {et~al.} 2011, \apjs, 192, 3,
  \dodoi{10.1088/0067-0049/192/1/3}

\bibitem[{{Paxton} {et~al.}(2013){Paxton}, {Cantiello}, {Arras}, {Bildsten},
  {Brown}, {Dotter}, {Mankovich}, {Montgomery}, {Stello}, {Timmes}, \&
  {Townsend}}]{paxton2013}
{Paxton}, B., {Cantiello}, M., {Arras}, P., {et~al.} 2013, \apjs, 208, 4,
  \dodoi{10.1088/0067-0049/208/1/4}

\bibitem[{{Paxton} {et~al.}(2015){Paxton}, {Marchant}, {Schwab}, {Bauer},
  {Bildsten}, {Cantiello}, {Dessart}, {Farmer}, {Hu}, {Langer}, {Townsend},
  {Townsley}, \& {Timmes}}]{paxton2015}
{Paxton}, B., {Marchant}, P., {Schwab}, J., {et~al.} 2015, \apjs, 220, 15,
  \dodoi{10.1088/0067-0049/220/1/15}

\bibitem[{{Paxton} {et~al.}(2018){Paxton}, {Schwab}, {Bauer}, {Bildsten},
  {Blinnikov}, {Duffell}, {Farmer}, {Goldberg}, {Marchant}, {Sorokina},
  {Thoul}, {Townsend}, \& {Timmes}}]{paxton2018}
{Paxton}, B., {Schwab}, J., {Bauer}, E.~B., {et~al.} 2018, \apjs, 234, 34,
  \dodoi{10.3847/1538-4365/aaa5a8}

\bibitem[{{Paxton} {et~al.}(2019){Paxton}, {Smolec}, {Schwab}, {Gautschy},
  {Bildsten}, {Cantiello}, {Dotter}, {Farmer}, {Goldberg}, {Jermyn}, {Kanbur},
  {Marchant}, {Thoul}, {Townsend}, {Wolf}, {Zhang}, \& {Timmes}}]{paxton2019}
{Paxton}, B., {Smolec}, R., {Schwab}, J., {et~al.} 2019, \apjs, 243, 10,
  \dodoi{10.3847/1538-4365/ab2241}

\bibitem[{{Peebles}(1993)}]{peebles1993}
{Peebles}, P.~J.~E. 1993, {Principles of Physical Cosmology},
  \dodoi{10.1515/9780691206721}

\bibitem[{{Perlmutter} {et~al.}(1999){Perlmutter}, {Aldering}, {Goldhaber},
  {Knop}, {Nugent}, {Castro}, {Deustua}, {Fabbro}, {Goobar}, {Groom}, {Hook},
  {Kim}, {Kim}, {Lee}, {Nunes}, {Pain}, {Pennypacker}, {Quimby}, {Lidman},
  {Ellis}, {Irwin}, {McMahon}, {Ruiz-Lapuente}, {Walton}, {Schaefer}, {Boyle},
  {Filippenko}, {Matheson}, {Fruchter}, {Panagia}, {Newberg}, {Couch}, \&
  {Project}}]{perlmutter1999}
{Perlmutter}, S., {Aldering}, G., {Goldhaber}, G., {et~al.} 1999, \apj, 517,
  565, \dodoi{10.1086/307221}

\bibitem[{{Planck Collaboration} {et~al.}(2020){Planck Collaboration},
  {Aghanim}, {Akrami}, {Ashdown}, {Aumont}, {Baccigalupi}, {Ballardini},
  {Banday}, {Barreiro}, {Bartolo}, {Basak}, {Battye}, {Benabed}, {Bernard},
  {Bersanelli}, {Bielewicz}, {Bock}, {Bond}, {Borrill}, {Bouchet}, {Boulanger},
  {Bucher}, {Burigana}, {Butler}, {Calabrese}, {Cardoso}, {Carron},
  {Challinor}, {Chiang}, {Chluba}, {Colombo}, {Combet}, {Contreras}, {Crill},
  {Cuttaia}, {de Bernardis}, {de Zotti}, {Delabrouille}, {Delouis}, {Di
  Valentino}, {Diego}, {Dor{\'e}}, {Douspis}, {Ducout}, {Dupac}, {Dusini},
  {Efstathiou}, {Elsner}, {En{\ss}lin}, {Eriksen}, {Fantaye}, {Farhang},
  {Fergusson}, {Fernandez-Cobos}, {Finelli}, {Forastieri}, {Frailis},
  {Fraisse}, {Franceschi}, {Frolov}, {Galeotta}, {Galli}, {Ganga},
  {G{\'e}nova-Santos}, {Gerbino}, {Ghosh}, {Gonz{\'a}lez-Nuevo}, {G{\'o}rski},
  {Gratton}, {Gruppuso}, {Gudmundsson}, {Hamann}, {Handley}, {Hansen},
  {Herranz}, {Hildebrandt}, {Hivon}, {Huang}, {Jaffe}, {Jones}, {Karakci},
  {Keih{\"a}nen}, {Keskitalo}, {Kiiveri}, {Kim}, {Kisner}, {Knox},
  {Krachmalnicoff}, {Kunz}, {Kurki-Suonio}, {Lagache}, {Lamarre}, {Lasenby},
  {Lattanzi}, {Lawrence}, {Le Jeune}, {Lemos}, {Lesgourgues}, {Levrier},
  {Lewis}, {Liguori}, {Lilje}, {Lilley}, {Lindholm}, {L{\'o}pez-Caniego},
  {Lubin}, {Ma}, {Mac{\'\i}as-P{\'e}rez}, {Maggio}, {Maino}, {Mandolesi},
  {Mangilli}, {Marcos-Caballero}, {Maris}, {Martin}, {Martinelli},
  {Mart{\'\i}nez-Gonz{\'a}lez}, {Matarrese}, {Mauri}, {McEwen}, {Meinhold},
  {Melchiorri}, {Mennella}, {Migliaccio}, {Millea}, {Mitra},
  {Miville-Desch{\^e}nes}, {Molinari}, {Montier}, {Morgante}, {Moss}, {Natoli},
  {N{\o}rgaard-Nielsen}, {Pagano}, {Paoletti}, {Partridge}, {Patanchon},
  {Peiris}, {Perrotta}, {Pettorino}, {Piacentini}, {Polastri}, {Polenta},
  {Puget}, {Rachen}, {Reinecke}, {Remazeilles}, {Renzi}, {Rocha}, {Rosset},
  {Roudier}, {Rubi{\~n}o-Mart{\'\i}n}, {Ruiz-Granados}, {Salvati}, {Sandri},
  {Savelainen}, {Scott}, {Shellard}, {Sirignano}, {Sirri}, {Spencer},
  {Sunyaev}, {Suur-Uski}, {Tauber}, {Tavagnacco}, {Tenti}, {Toffolatti},
  {Tomasi}, {Trombetti}, {Valenziano}, {Valiviita}, {Van Tent}, {Vibert},
  {Vielva}, {Villa}, {Vittorio}, {Wandelt}, {Wehus}, {White}, {White},
  {Zacchei}, \& {Zonca}}]{Planck2020}
{Planck Collaboration}, {Aghanim}, N., {Akrami}, Y., {et~al.} 2020, \aap, 641,
  A6, \dodoi{10.1051/0004-6361/201833910}

\bibitem[{{Popova} {et~al.}(1982){Popova}, {Tutukov}, \&
  {Yungelson}}]{popova1982}
{Popova}, E.~I., {Tutukov}, A.~V., \& {Yungelson}, L.~R. 1982, \apss, 88, 55,
  \dodoi{10.1007/BF00648989}

\bibitem[{{Prole} {et~al.}(2022){Prole}, {Clark}, {Klessen}, \&
  {Glover}}]{prole2022}
{Prole}, L.~R., {Clark}, P.~C., {Klessen}, R.~S., \& {Glover}, S. C.~O. 2022,
  \mnras, 510, 4019, \dodoi{10.1093/mnras/stab3697}

\bibitem[{{Punturo} {et~al.}(2010){Punturo}, {Abernathy}, {Acernese}, {Allen},
  {Andersson}, {Arun}, {Barone}, {Barr}, {Barsuglia}, {Beker}, {Beveridge},
  {Birindelli}, {Bose}, {Bosi}, {Braccini}, {Bradaschia}, {Bulik}, {Calloni},
  {Cella}, {Chassande Mottin}, {Chelkowski}, {Chincarini}, {Clark}, {Coccia},
  {Colacino}, {Colas}, {Cumming}, {Cunningham}, {Cuoco}, {Danilishin},
  {Danzmann}, {De Luca}, {De Salvo}, {Dent}, {De Rosa}, {Di Fiore}, {Di
  Virgilio}, {Doets}, {Fafone}, {Falferi}, {Flaminio}, {Franc}, {Frasconi},
  {Freise}, {Fulda}, {Gair}, {Gemme}, {Gennai}, {Giazotto}, {Glampedakis},
  {Granata}, {Grote}, {Guidi}, {Hammond}, {Hannam}, {Harms}, {Heinert},
  {Hendry}, {Heng}, {Hennes}, {Hild}, {Hough}, {Husa}, {Huttner}, {Jones},
  {Khalili}, {Kokeyama}, {Kokkotas}, {Krishnan}, {Lorenzini}, {L{\"u}ck},
  {Majorana}, {Mandel}, {Mandic}, {Martin}, {Michel}, {Minenkov}, {Morgado},
  {Mosca}, {Mours}, {M{\"u}ller{\textendash}Ebhardt}, {Murray}, {Nawrodt},
  {Nelson}, {Oshaughnessy}, {Ott}, {Palomba}, {Paoli}, {Parguez},
  {Pasqualetti}, {Passaquieti}, {Passuello}, {Pinard}, {Poggiani}, {Popolizio},
  {Prato}, {Puppo}, {Rabeling}, {Rapagnani}, {Read}, {Regimbau}, {Rehbein},
  {Reid}, {Rezzolla}, {Ricci}, {Richard}, {Rocchi}, {Rowan}, {R{\"u}diger},
  {Sassolas}, {Sathyaprakash}, {Schnabel}, {Schwarz}, {Seidel}, {Sintes},
  {Somiya}, {Speirits}, {Strain}, {Strigin}, {Sutton}, {Tarabrin},
  {Th{\"u}ring}, {van den Brand}, {van Leewen}, {van Veggel}, {van den Broeck},
  {Vecchio}, {Veitch}, {Vetrano}, {Vicere}, {Vyatchanin}, {Willke}, {Woan},
  {Wolfango}, \& {Yamamoto}}]{punturo2010}
{Punturo}, M., {Abernathy}, M., {Acernese}, F., {et~al.} 2010, Classical and
  Quantum Gravity, 27, 194002, \dodoi{10.1088/0264-9381/27/19/194002}

\bibitem[{{Reg{\H{o}}s} \& {Vink{\'o}}(2019)}]{regos2019}
{Reg{\H{o}}s}, E., \& {Vink{\'o}}, J. 2019, \apj, 874, 158,
  \dodoi{10.3847/1538-4357/ab0a73}

\bibitem[{{Reitze} {et~al.}(2019){Reitze}, {Adhikari}, {Ballmer}, {Barish},
  {Barsotti}, {Billingsley}, {Brown}, {Chen}, {Coyne}, {Eisenstein}, {Evans},
  {Fritschel}, {Hall}, {Lazzarini}, {Lovelace}, {Read}, {Sathyaprakash},
  {Shoemaker}, {Smith}, {Torrie}, {Vitale}, {Weiss}, {Wipf}, \&
  {Zucker}}]{reitze2019}
{Reitze}, D., {Adhikari}, R.~X., {Ballmer}, S., {et~al.} 2019, in Bulletin of
  the American Astronomical Society, Vol.~51, 35,
  \dodoi{10.48550/arXiv.1907.04833}

\bibitem[{{Riess} {et~al.}(1998){Riess}, {Filippenko}, {Challis},
  {Clocchiatti}, {Diercks}, {Garnavich}, {Gilliland}, {Hogan}, {Jha},
  {Kirshner}, {Leibundgut}, {Phillips}, {Reiss}, {Schmidt}, {Schommer},
  {Smith}, {Spyromilio}, {Stubbs}, {Suntzeff}, \& {Tonry}}]{riess1998}
{Riess}, A.~G., {Filippenko}, A.~V., {Challis}, P., {et~al.} 1998, \aj, 116,
  1009, \dodoi{10.1086/300499}

\bibitem[{{Rodney} {et~al.}(2014){Rodney}, {Riess}, {Strolger}, {Dahlen},
  {Graur}, {Casertano}, {Dickinson}, {Ferguson}, {Garnavich}, {Hayden}, {Jha},
  {Jones}, {Kirshner}, {Koekemoer}, {McCully}, {Mobasher}, {Patel}, {Weiner},
  {Cenko}, {Clubb}, {Cooper}, {Filippenko}, {Frederiksen}, {Hjorth},
  {Leibundgut}, {Matheson}, {Nayyeri}, {Penner}, {Trump}, {Silverman}, {U},
  {Azalee Bostroem}, {Challis}, {Rajan}, {Wolff}, {Faber}, {Grogin}, \&
  {Kocevski}}]{rodney2014}
{Rodney}, S.~A., {Riess}, A.~G., {Strolger}, L.-G., {et~al.} 2014, \aj, 148,
  13, \dodoi{10.1088/0004-6256/148/1/13}

\bibitem[{{Ruiter} {et~al.}(2009){Ruiter}, {Belczynski}, \&
  {Fryer}}]{ruiter2009}
{Ruiter}, A.~J., {Belczynski}, K., \& {Fryer}, C. 2009, \apj, 699, 2026,
  \dodoi{10.1088/0004-637X/699/2/2026}

\bibitem[{{Ruiz-Lapuente} {et~al.}(2018){Ruiz-Lapuente}, {Damiani}, {Bedin},
  {Gonz{\'a}lez Hern{\'a}ndez}, {Galbany}, {Pritchard}, {Canal}, \&
  {M{\'e}ndez}}]{ruiz2018}
{Ruiz-Lapuente}, P., {Damiani}, F., {Bedin}, L., {et~al.} 2018, \apj, 862, 124,
  \dodoi{10.3847/1538-4357/aac9c4}

\bibitem[{{Sana} {et~al.}(2012){Sana}, {de Mink}, {de Koter}, {Langer},
  {Evans}, {Gieles}, {Gosset}, {Izzard}, {Le Bouquin}, \&
  {Schneider}}]{sana2012}
{Sana}, H., {de Mink}, S.~E., {de Koter}, A., {et~al.} 2012, Science, 337, 444,
  \dodoi{10.1126/science.1223344}

\bibitem[{{Santoliquido} {et~al.}(2020){Santoliquido}, {Mapelli}, {Bouffanais},
  {Giacobbo}, {Di Carlo}, {Rastello}, {Artale}, \&
  {Ballone}}]{Santoliquido2020}
{Santoliquido}, F., {Mapelli}, M., {Bouffanais}, Y., {et~al.} 2020, \apj, 898,
  152, \dodoi{10.3847/1538-4357/ab9b78}

\bibitem[{{Santoliquido} {et~al.}(2021){Santoliquido}, {Mapelli}, {Giacobbo},
  {Bouffanais}, \& {Artale}}]{Santoliquido2021}
{Santoliquido}, F., {Mapelli}, M., {Giacobbo}, N., {Bouffanais}, Y., \&
  {Artale}, M.~C. 2021, \mnras, 502, 4877, \dodoi{10.1093/mnras/stab280}

\bibitem[{{Santoliquido} {et~al.}(2023){Santoliquido}, {Mapelli}, {Iorio},
  {Costa}, {Glover}, {Hartwig}, {Klessen}, \& {Merli}}]{santoliquido2023}
{Santoliquido}, F., {Mapelli}, M., {Iorio}, G., {et~al.} 2023, \mnras, 524,
  307, \dodoi{10.1093/mnras/stad1860}

\bibitem[{{Sarmento} {et~al.}(2018){Sarmento}, {Scannapieco}, \&
  {Cohen}}]{sarmento2018}
{Sarmento}, R., {Scannapieco}, E., \& {Cohen}, S. 2018, \apj, 854, 75,
  \dodoi{10.3847/1538-4357/aa989a}

\bibitem[{{Sarmento} {et~al.}(2019){Sarmento}, {Scannapieco}, \&
  {C{\^o}t{\'e}}}]{sarmento2019}
{Sarmento}, R., {Scannapieco}, E., \& {C{\^o}t{\'e}}, B. 2019, \apj, 871, 206,
  \dodoi{10.3847/1538-4357/aafa1a}

\bibitem[{{Scannapieco} {et~al.}(2005){Scannapieco}, {Madau}, {Woosley},
  {Heger}, \& {Ferrara}}]{scanapieco2005}
{Scannapieco}, E., {Madau}, P., {Woosley}, S., {Heger}, A., \& {Ferrara}, A.
  2005, \apj, 633, 1031, \dodoi{10.1086/444450}

\bibitem[{{Schaerer}(2002)}]{schaerer2002}
{Schaerer}, D. 2002, \aap, 382, 28, \dodoi{10.1051/0004-6361:20011619}

\bibitem[{{Schmidt} {et~al.}(1998){Schmidt}, {Suntzeff}, {Phillips},
  {Schommer}, {Clocchiatti}, {Kirshner}, {Garnavich}, {Challis}, {Leibundgut},
  {Spyromilio}, {Riess}, {Filippenko}, {Hamuy}, {Smith}, {Hogan}, {Stubbs},
  {Diercks}, {Reiss}, {Gilliland}, {Tonry}, {Maza}, {Dressler}, {Walsh}, \&
  {Ciardullo}}]{schmidt1998}
{Schmidt}, B.~P., {Suntzeff}, N.~B., {Phillips}, M.~M., {et~al.} 1998, \apj,
  507, 46, \dodoi{10.1086/306308}

\bibitem[{{Scolnic} {et~al.}(2018){Scolnic}, {Jones}, {Rest}, {Pan},
  {Chornock}, {Foley}, {Huber}, {Kessler}, {Narayan}, {Riess}, {Rodney},
  {Berger}, {Brout}, {Challis}, {Drout}, {Finkbeiner}, {Lunnan}, {Kirshner},
  {Sanders}, {Schlafly}, {Smartt}, {Stubbs}, {Tonry}, {Wood-Vasey}, {Foley},
  {Hand}, {Johnson}, {Burgett}, {Chambers}, {Draper}, {Hodapp}, {Kaiser},
  {Kudritzki}, {Magnier}, {Metcalfe}, {Bresolin}, {Gall}, {Kotak}, {McCrum}, \&
  {Smith}}]{scolnic2018}
{Scolnic}, D.~M., {Jones}, D.~O., {Rest}, A., {et~al.} 2018, \apj, 859, 101,
  \dodoi{10.3847/1538-4357/aab9bb}

\bibitem[{{Shao} \& {Li}(2021)}]{Shaoy2021}
{Shao}, Y., \& {Li}, X.-D. 2021, \apj, 908, 67,
  \dodoi{10.3847/1538-4357/abd2b4}

\bibitem[{{Sharda} {et~al.}(2020){Sharda}, {Federrath}, \&
  {Krumholz}}]{sharda2020}
{Sharda}, P., {Federrath}, C., \& {Krumholz}, M.~R. 2020, \mnras, 497, 336,
  \dodoi{10.1093/mnras/staa1926}

\bibitem[{{Shivvers} {et~al.}(2017){Shivvers}, {Modjaz}, {Zheng}, {Liu},
  {Filippenko}, {Silverman}, {Matheson}, {Pastorello}, {Graur}, {Foley},
  {Chornock}, {Smith}, {Leaman}, \& {Benetti}}]{shivvers2017}
{Shivvers}, I., {Modjaz}, M., {Zheng}, W., {et~al.} 2017, \pasp, 129, 054201,
  \dodoi{10.1088/1538-3873/aa54a6}

\bibitem[{{Skinner} \& {Wise}(2020)}]{skinner2020}
{Skinner}, D., \& {Wise}, J.~H. 2020, \mnras, 492, 4386,
  \dodoi{10.1093/mnras/staa139}

\bibitem[{{Song} {et~al.}(2020){Song}, {Meynet}, {Li}, {Peng}, {Zhang}, \&
  {Zhan}}]{songh2020}
{Song}, H., {Meynet}, G., {Li}, Z., {et~al.} 2020, \apj, 892, 41,
  \dodoi{10.3847/1538-4357/ab7993}

\bibitem[{{Stacy} \& {Bromm}(2013)}]{stacy2013}
{Stacy}, A., \& {Bromm}, V. 2013, \mnras, 433, 1094,
  \dodoi{10.1093/mnras/stt789}

\bibitem[{{Stacy} {et~al.}(2016){Stacy}, {Bromm}, \& {Lee}}]{stacy2016}
{Stacy}, A., {Bromm}, V., \& {Lee}, A.~T. 2016, \mnras, 462, 1307,
  \dodoi{10.1093/mnras/stw1728}

\bibitem[{{Strolger} {et~al.}(2020){Strolger}, {Rodney}, {Pacifici}, {Narayan},
  \& {Graur}}]{strolger2020}
{Strolger}, L.-G., {Rodney}, S.~A., {Pacifici}, C., {Narayan}, G., \& {Graur},
  O. 2020, \apj, 890, 140, \dodoi{10.3847/1538-4357/ab6a97}

\bibitem[{{Susa} {et~al.}(2014){Susa}, {Hasegawa}, \& {Tominaga}}]{susa2014}
{Susa}, H., {Hasegawa}, K., \& {Tominaga}, N. 2014, \apj, 792, 32,
  \dodoi{10.1088/0004-637X/792/1/32}

\bibitem[{{Tanaka} {et~al.}(2013){Tanaka}, {Moriya}, \& {Yoshida}}]{tanaka2013}
{Tanaka}, M., {Moriya}, T.~J., \& {Yoshida}, N. 2013, \mnras, 435, 2483,
  \dodoi{10.1093/mnras/stt1469}

\bibitem[{{Toonen} {et~al.}(2012){Toonen}, {Nelemans}, \& {Portegies
  Zwart}}]{toonen2012}
{Toonen}, S., {Nelemans}, G., \& {Portegies Zwart}, S. 2012, \aap, 546, A70,
  \dodoi{10.1051/0004-6361/201218966}

\bibitem[{{Trenti} \& {Stiavelli}(2009)}]{trenti2009}
{Trenti}, M., \& {Stiavelli}, M. 2009, \apj, 694, 879,
  \dodoi{10.1088/0004-637X/694/2/879}

\bibitem[{{Tsai} {et~al.}(2023){Tsai}, {Chen}, {Whalen}, {Ou}, \&
  {Woods}}]{tsai2023}
{Tsai}, S.-H., {Chen}, K.-J., {Whalen}, D., {Ou}, P.-S., \& {Woods}, T.~E.
  2023, \apj, 951, 84, \dodoi{10.3847/1538-4357/acd936}

\bibitem[{{Turk} {et~al.}(2009){Turk}, {Abel}, \& {O'Shea}}]{turk2009}
{Turk}, M.~J., {Abel}, T., \& {O'Shea}, B. 2009, Science, 325, 601,
  \dodoi{10.1126/science.1173540}

\bibitem[{{Visbal} {et~al.}(2020){Visbal}, {Bryan}, \& {Haiman}}]{visbal2020}
{Visbal}, E., {Bryan}, G.~L., \& {Haiman}, Z. 2020, \apj, 897, 95,
  \dodoi{10.3847/1538-4357/ab994e}

\bibitem[{{Wang} \& {Han}(2012)}]{wangb2012}
{Wang}, B., \& {Han}, Z. 2012, \nar, 56, 122,
  \dodoi{10.1016/j.newar.2012.04.001}

\bibitem[{{Wang} {et~al.}(2017){Wang}, {Baade}, {Baron}, {Bernard}, {Bromm},
  {Brown}, {Clayton}, {Cooke}, {Croton}, {Curtin}, {Drout}, {Doi}, {Dominguez},
  {Finkelstein}, {Gal-Yam}, {Geil}, {Heger}, {Hoeflich}, {Jian}, {Krisciunas},
  {Koekemoer}, {Lunnan}, {Maeda}, {Maund}, {Modjaz}, {Mould}, {Nomoto},
  {Nugent}, {Patat}, {Pacucci}, {Phillips}, {Rest}, {Regos}, {Sand}, {Sparks},
  {Spyromilio}, {Staveley-Smith}, {Suntzeff}, {Uddin}, {Villarroel}, {Vinko},
  {Whalen}, {Wheeler}, {Wood-Vasey}, {Yang}, \& {Yue}}]{wangl2017}
{Wang}, L., {Baade}, D., {Baron}, E., {et~al.} 2017, arXiv e-prints,
  arXiv:1710.07005, \dodoi{10.48550/arXiv.1710.07005}

\bibitem[{{Wang} {et~al.}(2022){Wang}, {Cheng}, {Ge}, {Meng}, {Daddi}, {Yan},
  {Jones}, {Malkan}, {Arrabal Haro}, {Brammer}, \& {Oguri}}]{wangx2022}
{Wang}, X., {Cheng}, C., {Ge}, J., {et~al.} 2022, arXiv e-prints,
  arXiv:2212.04476, \dodoi{10.48550/arXiv.2212.04476}

\bibitem[{{Webbink}(1984)}]{webbink1984}
{Webbink}, R.~F. 1984, \apj, 277, 355, \dodoi{10.1086/161701}

\bibitem[{{Welch} {et~al.}(2022){Welch}, {Coe}, {Diego}, {Zitrin},
  {Zackrisson}, {Dimauro}, {Jim{\'e}nez-Teja}, {Kelly}, {Mahler}, {Oguri},
  {Timmes}, {Windhorst}, {Florian}, {de Mink}, {Avila}, {Anderson}, {Bradley},
  {Sharon}, {Vikaeus}, {McCandliss}, {Brada{\v{c}}}, {Rigby}, {Frye}, {Toft},
  {Strait}, {Trenti}, {Sharma}, {Andrade-Santos}, \& {Broadhurst}}]{welch2022}
{Welch}, B., {Coe}, D., {Diego}, J.~M., {et~al.} 2022, \nat, 603, 815,
  \dodoi{10.1038/s41586-022-04449-y}

\bibitem[{{Whalen} {et~al.}(2013{\natexlab{a}}){Whalen}, {Fryer}, {Holz},
  {Heger}, {Woosley}, {Stiavelli}, {Even}, \& {Frey}}]{whalen2013c}
{Whalen}, D.~J., {Fryer}, C.~L., {Holz}, D.~E., {et~al.} 2013{\natexlab{a}},
  \apjl, 762, L6, \dodoi{10.1088/2041-8205/762/1/L6}

\bibitem[{{Whalen} {et~al.}(2013{\natexlab{b}}){Whalen}, {Joggerst}, {Fryer},
  {Stiavelli}, {Heger}, \& {Holz}}]{whalen2013b}
{Whalen}, D.~J., {Joggerst}, C.~C., {Fryer}, C.~L., {et~al.}
  2013{\natexlab{b}}, \apj, 768, 95, \dodoi{10.1088/0004-637X/768/1/95}

\bibitem[{{Whalen} {et~al.}(2014{\natexlab{a}}){Whalen}, {Smidt}, {Even},
  {Woosley}, {Heger}, {Stiavelli}, \& {Fryer}}]{whalen2014a}
{Whalen}, D.~J., {Smidt}, J., {Even}, W., {et~al.} 2014{\natexlab{a}}, \apj,
  781, 106, \dodoi{10.1088/0004-637X/781/2/106}

\bibitem[{{Whalen} {et~al.}(2014{\natexlab{b}}){Whalen}, {Smidt}, {Heger},
  {Hirschi}, {Yusof}, {Even}, {Fryer}, {Stiavelli}, {Chen}, \&
  {Joggerst}}]{whalen2014b}
{Whalen}, D.~J., {Smidt}, J., {Heger}, A., {et~al.} 2014{\natexlab{b}}, \apj,
  797, 9, \dodoi{10.1088/0004-637X/797/1/9}

\bibitem[{{Whelan} \& {Iben}(1973)}]{whelan1973}
{Whelan}, J., \& {Iben}, Icko, J. 1973, \apj, 186, 1007, \dodoi{10.1086/152565}

\bibitem[{{Willems} \& {Kolb}(2004)}]{willems2004}
{Willems}, B., \& {Kolb}, U. 2004, \aap, 419, 1057,
  \dodoi{10.1051/0004-6361:20040085}

\bibitem[{{Wollenberg} {et~al.}(2020){Wollenberg}, {Glover}, {Clark}, \&
  {Klessen}}]{wollenberg2020}
{Wollenberg}, K. M.~J., {Glover}, S. C.~O., {Clark}, P.~C., \& {Klessen}, R.~S.
  2020, \mnras, 494, 1871, \dodoi{10.1093/mnras/staa289}

\bibitem[{{Yan} {et~al.}(2023){Yan}, {Ma}, {Sun}, {Wang}, {Kelly}, {Diego},
  {Cohen}, {Windhorst}, {Jansen}, {Grogin}, {Beacom}, {Conselice}, {Driver},
  {Frye}, {Coe}, {Marshall}, {Koekemoer}, {Willmer}, {Robotham}, {D'Silva},
  {Summers}, {Nonino}, {Pirzkal}, {Ryan}, {Ortiz}, {Tompkins}, {Bhatawdekar},
  {Cheng}, {Zitrin}, \& {Willner}}]{yanh2023}
{Yan}, H., {Ma}, Z., {Sun}, B., {et~al.} 2023, \apjs, 269, 43,
  \dodoi{10.3847/1538-4365/ad0298}

\bibitem[{{Yoshida} {et~al.}(2008){Yoshida}, {Omukai}, \&
  {Hernquist}}]{yoshida2008}
{Yoshida}, N., {Omukai}, K., \& {Hernquist}, L. 2008, Science, 321, 669,
  \dodoi{10.1126/science.1160259}

\end{thebibliography}

\end{document}